\documentclass[11pt]{article}
\usepackage{subfigure}
\usepackage{graphicx}
\usepackage{amsmath,amsfonts}
\usepackage{algorithm,algorithmic}
\usepackage{listings}
\usepackage{natbib}

\DeclareMathOperator{\erf}{erf}

\newcommand{\model}{{\cal M}}
\newcommand{\data}{{\bf D}}
\newcommand{\param}{\boldsymbol{\theta}}
\newcommand{\bx}{\boldsymbol{x}}
\newcommand{\barm}{{\bar m}}

\begin{document}

\title{\vspace{-3cm}
  Computing the Bayes Factor from a Markov chain Monte Carlo
  Simulation of the Posterior Distribution}

\author{Martin D. Weinberg\thanks{E-mail:
    weinberg@astro.umass.edu}\\Department of Astronomy\\University of
  Massachusetts, Amherst, USA}

\date{}

\maketitle

\begin{abstract}
  Computation of the marginal likelihood from a simulated posterior
  distribution is central to Bayesian model selection but is
  computationally difficult.  The often-used harmonic mean
  approximation uses the posterior directly but is unstably sensitive
  to samples with anomalously small values of the likelihood.  The
  Laplace approximation is stable but makes strong, and often
  inappropriate, assumptions about the shape of the posterior
  distribution.  It is useful, but not general.  We need algorithms
  that apply to general distributions, like the harmonic mean
  approximation, but do not suffer from convergence and instability
  issues.  Here, I argue that the marginal likelihood can be reliably
  computed from a posterior sample by careful attention to the
  numerics of the probability integral.  Posing the expression for the
  marginal likelihood as a Lebesgue integral, we may convert the
  harmonic mean approximation from a sample statistic to a quadrature
  rule.  As a quadrature, the harmonic mean approximation suffers from
  enormous truncation error as consequence .  This error is a direct
  consequence of poor coverage of the sample space; the posterior
  sample required for accurate computation of the marginal likelihood
  is much larger than that required to characterize the posterior
  distribution when using the harmonic mean approximation. In
  addition, I demonstrate that the integral expression for the
  harmonic-mean approximation converges slowly at best for
  high-dimensional problems with uninformative prior
  distributions. These observations lead to two computationally-modest
  families of quadrature algorithms that use the full generality
  sample posterior but without the instability.  The first algorithm
  automatically eliminates the part of the sample that contributes
  large truncation error.  The second algorithm uses the posterior
  sample to assign probability to a partition of the sample space and
  performs the marginal likelihood integral directly.  This eliminates
  convergence issues. The first algorithm is analogous to standard
  quadrature but can only be applied for convergent problems.  The
  second is a hybrid of cubature: it uses the posterior to discover
  and tessellate the subset of that sample space was explored and uses
  quantiles to compute a representative field value.  Qualitatively,
  the first algorithm improves the harmonic mean approximation using
  numerical analysis, and the second algorithm is an adaptive version
  of the Laplace approximation.  Neither algorithm makes strong
  assumptions about the shape of the posterior distribution and
  neither is sensitive to outliers.  Based on numerical tests, we
  recommend a combined application of both algorithms as consistency
  check to achieve a reliable estimate of the marginal likelihood from
  a simulated posterior distribution.  \par\bigskip\noindent {\bf
    Keywords:} Bayesian computation, marginal likelihood, algorithm,
  Bayes factors, model selection
\end{abstract}

\section{Introduction}
\label{sec:intro}

A Bayesian data analysis specifies joint probability distributions to
describe the relationship between the prior information, the model or
hypotheses, and the data.  Using Bayes theorem, the posterior
distribution is uniquely determined from the conditional probability
distribution of the unknowns given the observed data.  The posterior
probability is usually stated as follows:
\begin{equation}
  P(\param|\model,\data) = \frac{\pi(\param|\model)L(\data|\param,\model)}{Z}
  \label{eq:bayes}
\end{equation}
where
\begin{equation}
  Z \equiv P(\data|\model) = \int d\param\,\pi(\param|\model)
  L(\data|\param,\model)
  \label{eq:evidence}
\end{equation}
is the marginal likelihood.  The symbol ${\cal M}$ denotes the
assumption of a particular model and the parameter vector
$\param\in\Omega$.  For physical models, the sample space $\Omega$ is
most often a continuous space.  In words, equation (\ref{eq:bayes})
says: the probability of the model parameters given the data and the
model is proportional to the prior probability of the model parameters
and the probability of the data given the model parameters.  The
posterior may be used, for example, to infer the distribution of model
parameters or to discriminate between competing hypotheses or models.
The latter is particularly valuable given the wide variety of
astronomical problems where diverse hypotheses describing
heterogeneous physical systems is the norm \citep[see][for a thorough
discussion of Bayesian data analysis]{Gelman.etal:03}.

For parameter estimation, one often considers $P(\data|\model)$ to be
an uninteresting normalization constant.  However, equation
(\ref{eq:evidence}) clearly admits a meaningful interpretation: it is
the support or \emph{evidence} for a model given the data.  This see
this, assume that the prior probability of some model, $\model_j$ say,
is $P(\model_j)$.  Then by Bayes theorem, the probability of the model
given the data is $P(\model_j|\data) = P(\model_j) P(\data|\model_j) /
P(\data)$.  The posterior odds of Model $j=0$ relative to Model $j=1$
is then:
\begin{equation*}
  \frac{P(\model_0|\data)}{P(\model_1|\data)} =
  \frac{P(\model_0)}{P(\model_1)} \frac{P(\data|\model_0)
  }{P(\data|\model_1)}.
\end{equation*}
If we have information about the ratio of prior odds,
$P(\model_0)/P(\model_1)$, we should use it, but more often than not
our lack of knowledge forces a choice of $P(\model_0)/P(\model_1)=1$.
Then, we estimate the relative probability of the models given $\data$
over their prior odds by the Bayes factor
$P(\data|\model_0)/P(\data|\model_1)$ \citep[see][for a discussion of
additional concerns]{Lavine.Schervish:99}.  When there is no
ambiguity, we will omit the explicit dependence on $\model$ of the
prior distribution, likelihood function, and marginal likelihood for
notational convenience.

The Bayes factor has a number of attractive advantages for model
selection \citep{Kass.Raftery:95}: (1) it is a consistent selector;
that is, the ratio will increasingly favor the true model in the limit
of large data; (2) Bayes factors act as an Occam's razor, preferring
the simpler model if the ``fits'' are similar; and (3) Bayes factors
do not require the models to be nested in any way; that is, the models
and their parameters need not be equivalent in any limit.  There is a
catch: direct computation of the marginal likelihood
(eq. \ref{eq:evidence}) is intractable for most problems of interest.
However, recent advances in computing technology together with
developments in Markov chain Monte Carlo (MCMC) algorithms have the
promise to compute the posterior distribution for problems that have
been previously infeasible owing to dimensionality or complexity.  The
posterior distribution is central to Bayesian inference: it summarizes
all of our knowledge about the parameters of our model and is the
basis for all subsequent inference and prediction for a given problem.
For example, current astronomical datasets are very large, the
proposed models may be high-dimensional, and therefore, the posterior
sample is expensive to compute.  However, once obtained, the posterior
sample may be exploited for a wide variety of tasks.  Although
dimension-switching algorithms, such as reversible-jump MCMC
\citep{Green:95} incorporate model selection automatically without
need for Bayes factors, these simulations appear slow to converge for
some of our real-world applications.  Moreover, the marginal
likelihood may be used for an endless variety of tests, ex post facto.

\citet{Newton.Raftery:94} presented a formula for estimating $Z$ from
a posterior distribution of parameters.  They noted that a MCMC
simulation of the posterior selects values of $\param\in\Omega$
distributed as
\[
Z\times P(\param|\data) = L(\data|\param) \pi(\param)
\]
and, therefore,
\begin{equation}
Z\times \int_\Omega d\param\,\frac{P(\param|\data)}{L(\data|\param)} = \int_\Omega
d\param\,\pi(\param) = 1
\label{eq:Zdef0}
\end{equation}
or
\begin{equation}
\frac1Z = \int_\Omega d\param\,\frac{P(\param|\data)}{L(\data|\param)} =
E\left[\frac{1}{L(\param|\data)}\right]_{P(\param|\data)},
\label{eq:hmean}
\end{equation}
having suppressed the explicit dependence on $\model$ for notational
clarity.  This latter equation says that the marginal likelihood is
the harmonic mean of the likelihood with respect to the posterior
distribution.  It follows that the harmonic mean computed from a
sampled posterior distribution is an estimator for the marginal
likelihood, e.g.:
\begin{equation}
{\tilde Z} = \left[\frac1N\sum_{i=1}^N \frac{1}{L(\data|\theta_i)}\right]^{-1}.
\label{eq:hsamp}
\end{equation}
Unfortunately, this estimator is prone to domination by a few outlying
terms with abnormally small values of $L_j$ \citep[e.g. see][and
  references therein]{Raftery.etal:07}.  \citet{Wolpert:02} describes
convergence criteria for equation (\ref{eq:hsamp}) and
\citet{Chib.Jeliazkov:01} present augmented approaches with error
estimates.

Alternative approaches to computing the marginal likelihood from the
posterior distribution have been described at length by
\citet{Kass.Raftery:95}.  Of these, the Laplace approximation, which
approximates the posterior distribution by a multidimensional Gaussian
distribution and uses this approximation to compute equation
(\ref{eq:evidence}) directly, is the most widely used.  This seems to
be favored over equation (\ref{eq:harmonic}) because of the problem
with outliers and hence because of convergence and stability.  In many
cases, however, the Laplace approximation is far from adequate in two
ways. First, one must identify all the dominant modes, and second,
modes may not be well-represented by a multidimensional Gaussian
distribution for problems of practical interest, although many
promising improvements have been suggested
\citep[e.g.][]{DiCicio.etal:97}.  \citet{Trotta:07} explored the use
of the Savage-Dickey density ratio for cosmological model selection
\citep[see also][for a full review of the model selection problem for
cosmology]{Trotta:08}.

Finally, we may consider evaluation of equation (\ref{eq:evidence})
directly.  The MCMC simulation samples the posterior distribution by
design, and therefore, can be used to construct volume elements in
$k$-dimensional parameter space, $d\param$, e.g. when
$\Omega\subset\mathbb{R}^k$.  Although the volume will be sparsely
sampled in regions of relatively low likelihood, these same volumes
will make little contribution to equation (\ref{eq:evidence}).  The
often-used approach from computational geometry, Delaney
triangulation, maximizes the minimum angle of the facets and thereby
yields the ``roundest'' volumes.  Unfortunately, the standard
procedure scales as ${\cal O}(kN^2)$ for a sample of $N$ points.  This
can be reduced to ${\cal O}(N\log N + N^{k/2})$ using the flip
algorithm with iterative construction \citep{EdelsbrunnerShah:96} but
this scaling is prohibitive for large $N$ and $k$ typical of many
problems.  Rather, in this paper, we consider the less optimal but
tractable kd-tree for space partitioning.

In part, the difficulty in computing the marginal likelihood from the
sampled posterior has recently led \citet[``nesting
sampling'']{Skilling:06} to suggest an algorithm to simulate the
marginal likelihood rather than the posterior distribution.  This idea
has been adopted and extended by cosmological modelers
\citep{Mukherjee.etal:06,Feroz.Hobson:08}.  The core idea of nesting
sampling follows by rewriting equation (\ref{eq:Zdef0}) as a double
integral and swapping the order of integration, e.g. 
\begin{equation}
Z = \int_\Omega d\theta \pi(\theta) \int_0^{L(\data|\theta)} dy = 
\int_0^{\sup\{L(\data|\theta): \theta\in\Omega\}} dy 
\int_{L(\data|\theta)>y} d\theta \pi(\theta).
\label{eq:nest}
\end{equation}
The nested sampler is a Monte Carlo sampler for the likelihood
function $L$ with respect to the prior distrbution $\pi$ so that
$L>y$.  The generalization of the construction in equation
(\ref{eq:nest}) for general distributions and multiple dimensions is
the Lebesgue integral (see \S\ref{sec:intgr}).  Clearly, this
procedure has no problems with outliers with small values of
$L(\data|\param)$.  Of course, any algorithm implementing nested
sampling must still thoroughly sample the multidimensional posterior
distribution and so retains all of the intendant difficulties that
MCMC has been designed to solve.

In many ways, the derivation of the nested sampler bears a strong
resemblance to the derivation of the harmonic mean but without any
obvious numerical difficulty.  This led me to a careful study of
equations (\ref{eq:evidence}) and (\ref{eq:Zdef0}) to see if the
divergence for small value of likelihood could be addressed.  Indeed
they can, and the following sections describe two algorithms based on
each of these equations.  These new algorithms retain the main
advantage of the harmonic mean approximation (HMA): direct
incorporation of the sampled posterior without any assumption of a
functional form.  In this sense, they are fully and automatically
adaptive to any degree multimodality given a sufficiently large
sample.  We begin in \S\ref{sec:intgr} with a background discussion of
Lebesgue integration applied to probability distributions and Monte
Carlo (MC) estimates.  We apply this in \S\ref{sec:evid} to the
marginal likelihood computation. This development both illuminates the
arbitrariness in the HMA from the numerical standpoint and leads to an
improved approach outlined in \S\ref{sec:algo}.  In short, the
proposed approach is motivated by methods of numerical quadrature
rather than sample statistics.  Examples in \S\ref{sec:examples}
compare the application of the new algorithms to the HMA and the
Laplace approximation.  The overall results are discussed and
summarized in \S\ref{sec:summary}.

\section{Integration on a random mesh}
\label{sec:intgr}

\begin{figure}[t]
\centering
\includegraphics[width=0.5\textwidth]{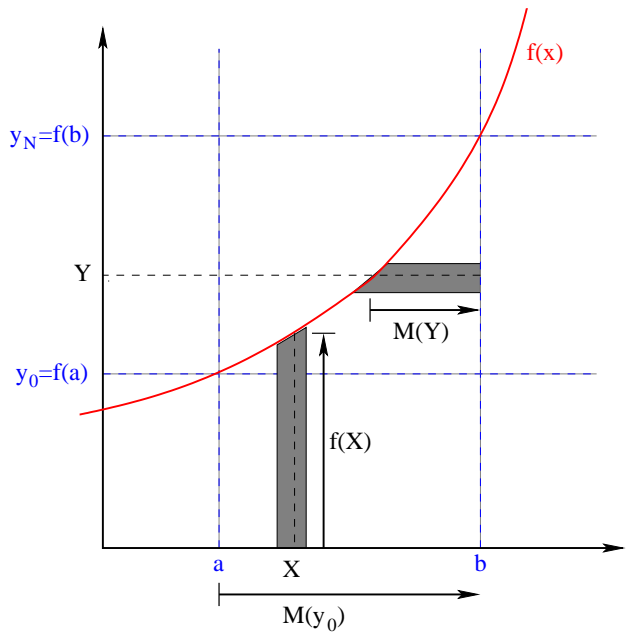}
\caption{Illustration of the integral $\int^b_adx\,f(x)$ using Riemann
  and Lebesgue integration.  For Riemann integration, we sum thin
  vertical rectangles of height $f(X)$ about the abscissa point $X$
  for some width $dx$.  For Lebesgue integration, we sum thin
  horizontal rectangles of width $M(Y)$ about the ordinate point $Y$ of
  height $dy$.  In both cases, we sum the area under the curve $f(x)$.
  In the Lebesgue case, we must add in the rectangle of width
  $M(y_0)$ and height $y_0$.}
\label{fig:geom}
\end{figure}

Assume that we have a MC generated sample of random variates with
density $f(x)$.  Recall that any moment or expectation with respect to
this density may be computed as a single sum, independent of the
parameter-space dimension!  This powerful and seemingly innocuous
property follows from the power of Lebesgue integration
\citep[e.g.][]{Capinski.Kopp:07}.  To see this, let us begin by
considering a one-dimensional integral
\begin{equation}
I = \int^b_a f(x)\,dx
\label{eq:int1}
\end{equation}
where $f(x)$ is non-negative, finite and bounded, that is: $0\le y_0
\le f(x) \le y_N$ for $x\in[a,b]$.  More formally, we may define the
Lebesgue integral $I$ over the measure of sets of points
$x\in\Omega=[a,b]$ with measure $\mu$ as follows.  We assume that
$f(x)$ is measurable over $\Omega$ and, following
\citet[\S8.3]{Temple:71}, define the measure function $M(y)=\mu(x|f(x)
> y)$.  Clearly, $M(y)$ is monotonic with $0=M(y_N)\le M(y)\le M(y_0)$
and $\mu(\Omega)=b-a$.  Let ${\cal S}: a \le x_0 < x_1 < \cdots < x_N
\le b$ be a partition of $[a, b]$ with
$\lim_{N\rightarrow\infty}x_0=a$ and $\lim_{N\rightarrow\infty}x_N=b$.
In our case, the partition ${\cal S}$ is our MC sample.  The choice of
${\cal S}$ induces a partition of $y$ through $y_j=f(x_j)$ although
the sequence in $y$ will no longer be monotonic.  For convenience,
reorder the $y$ such that $y_{i-1}\le y_{i}$.  Now consider the
Lebesgue integral of $f(x)$ over $\Omega$,
\begin{equation}
  I = \int^b_a f(x)\,dx = \int_{f^{-1}(y)\in\Omega} M(y)\,dy.
  \label{eq:int2}
\end{equation}
We interpret this geometrically as the area under the curve
$x=f^{-1}(y)$; in other words, we have swapped the abscissa and
ordinate.  To see this, define
\begin{equation}
  {\cal L}_{{\cal S}} = \sum_{i=1}^N M(y_{i-1})(y_{i} - y_{i-1})
  \quad\mbox{and}\quad
  {\cal U}_{{\cal S}} = \sum_{i=1}^N M(y_{i})(y_{i} - y_{i-1})
  \label{eq:LUn}
\end{equation}
for the partition ${\cal S}$.  Define the interval $\lambda_i\equiv
y_i - y_{i-1}$. Clearly
\begin{equation}
  {\cal U}_{{\cal S}} - {\cal L}_{{\cal S}} \le \sum_{i=1}^N
  \left[M(y_i) - M(y_{i-1})\right]\sup(\lambda_j) = \left[M(b) - M(a)\right]\sup(\lambda_j)
\end{equation}
and, therefore, $\lim_{N\rightarrow\infty}({\cal U}_{{\cal S}} - {\cal
  L}_{{\cal S}})\rightarrow0$ since $M(y)$ is monotonic and
$\lim_{N\rightarrow\infty}\lambda_j \rightarrow 0$.  Using this, we may
evaluate the integral in equation (\ref{eq:int2}) as follows:
\begin{equation}
I = M(y_0) y_0 + \lim_{N\rightarrow\infty}{\cal L}_{\cal S}
=   M(y_0) y_0 + \lim_{N\rightarrow\infty}{\cal U}_{\cal S}
=   M(y_0) y_0 + \int_{y_0}^{y_N} M(y)\,dy.
\label{eq:lebesgue2}
\end{equation}
The sums in equation (\ref{eq:LUn}) have the form of a rectangular
quadrature of the measure over the \emph{range} of $f(x)$.  This
geometry is illustrated in Figure \ref{fig:geom}.  Although the
Lebesgue integration theory is general, the equivalence of the
one-dimensional integral and equation (\ref{eq:lebesgue2}) is easily
seen by rewriting equation (\ref{eq:int1}) as a two dimensional
integral and changing the order of integration using elementary
techniques as follows:
\[
I = \int^b_a f(x)\,dx 
= \int^b_a \left\{\int_0^{f(x)} dy\right\}dx
=  M(y_0)y_0 + \int^{f_{max}}_{f_{min}} M(y) dy
\]
where $f_{min}$ and $f_{max}$ are the minimum and maximum values of
$f(x)$ in $[a, b]$.  

An average of the sums in equation (\ref{eq:LUn}) gives us a
trapezoidal rule analog:
\begin{equation}
  {\cal T}_{{\cal S}} = \frac12\sum_{i=1}^N \left[M(y_{i-1}) +
    M(y_{i})\right](y_{i} - y_{i-1}).
  \label{eq:Tn}
\end{equation}
Further generalization is supported by the Lebesgue theory of
differentiation.  A central result of the theory is that a continuous,
monotonic function in some interval is differentiable almost
everywhere in the interval \citep{Temple:71}.  This applies to the
measure function $M(Y)$.  This result may be intuitively motivated in
the context of our marginal likelihood calculation as follows.  Our
measure function describes the amount of density with likelihood
smaller than some particular value.  A typical likelihood surface for
a physical model is smooth, continuous, and typically consists of
several discrete modes.  Consider constructing $M(Y)$ by increasing
the value of $Y$ beginning at the point of maximum likelihood peak.
Since $Y=L^{-1}$, this is equivalent to beginning at
$\max(L)=Y^{-1}_0$ and decreasing $L$.  Recall that $M(Y)$ decreases
from 1 to 0 as $Y$ increases from $Y_0$ to $\infty$.  Therefore, we
construct $M(Y)$ from $M(Y+\Delta Y)$ by finding the level set
corresponding to some value of $Y$ and subtracting off the area of the
likelihood surface constructed from the perimeter of the set times
$\Delta Y$.  The function $M(Y)$ will decrease smoothly from unity at
$Y_0$ until $Y$ reaches the peak of the second mode.  At this point,
there may be a discontinuity in the derivative of $M(Y)$ as another
piece of perimeter joins the level set, but it will be smooth
thereafter until the peak of the third mode is reached, and so on.
Since we expect the contribution to $Z$ to be dominated by a few
primary modes, this suggests that we can evaluate the integral $I$
numerically using the quadrature implied by equation
(\ref{eq:lebesgue2}) and possibly even higher-order quadrature rules.
These arguments further suggest that partitioning $\Omega$ into
separate domains supporting individual modes would improve the
numerics by removing discontinuities in the derivative of $M(Y)$ and
explicitly permitting the use of higher-order quadrature rules. This
partition may be difficult to construct automatically, however.

To better control the truncation error for this quadrature, we might
like to choose a uniform partition in $y$, $\lambda=\lambda_i$, to
evaluate the sums ${\cal L}_N$ and ${\cal U}_N$.  For MC integration,
this is not possible.  Rather, MC selects ${\cal S}$ with irregular
spacings and this induces a partition of $y$.  Motivated by kernel
density estimation, we may then approximate $M(y)$ by
\begin{equation}
  {\tilde M}(y) = \frac1N\sum_{j=0}^N\Theta_j(y)
\label{eq:nummeas}
\end{equation}
where $\Theta(\cdot)$ monotonically increases from 0 to 1 in the
vicinity of $f(x_j)$.  For example, we may choose $\Theta$ to be the
Heaviside function
\begin{equation}
  \Theta_j(y) =
  \begin{cases}
    1 & y < f(x_j) \quad\mbox{or}\quad y \le f(x_j), \\
    0 & \mbox{otherwise}
  \end{cases}
\label{eq:nummeas2}
\end{equation}
which assigns a ``step'' to the upper value of the range in $y$ for
each $x_j$.  Alternatively, we may consider smoothing functions such
as
\begin{equation}
  \Theta_j(y) = \frac12\left[1 + \erf\left(\frac{y -
        f(x_j)}{\alpha_j}\right)\right]
  \label{eq:nummeas3}
\end{equation}
where $\erf(\cdot)$ denotes the error function.  Then, upon substituting
equation (\ref{eq:nummeas}) into equation (\ref{eq:lebesgue2}) for
${\cal U}_{\cal S}$, we get:
\begin{equation}
  {\tilde I}_N = M(y_0)y_0 + \sum_{j=1}^N M(y_{j-1})(y_j - y_{j-1}) 
  = \sum_{j=1}^N \mu_j f(x_j)
  \label{eq:ItildeN}
\end{equation}
where $\mu_j = {\tilde M}(y_{i-1}) - {\tilde M}(y_i)$ and ${\tilde
  M}(y_N)=0$ by construction and the final equality follows by
gathering common factors of $y_i = f(x_i)$ in the earlier summation.

For integration over probability distributions, we desire $x$
distributed according some probability density $g(x)$,
\begin{equation}
  I[f] = \int_{\mathbb{R}} f(x) g(x)\,dx,
  \label{eq:probint}
\end{equation}
which yields
\begin{equation}
  M(y) = \int_{f(x)>y} g(x)\,dx
\end{equation}
with the normalization $\int g(x)\,dx = 1$, and therefore,
$M(y)\in[0,1]$.  Despite the appearance
of the density $g$ in equation (\ref{eq:probint}), only the value of
measure has changed, not the formalism, i.e. the points $x_j$ are now
sampled from $g(x)$ rather than uniformly.

Let us now consider the connection between equation (\ref{eq:ItildeN})
and a classic MC integration, although we do not need this for later
sections of this paper.  For a classic MC integration, we choose
$\mu_j=\mu(\Omega)/N=\mbox{constant}$ by construction, and with this
substitution ${\tilde I}$ becomes the MC approximation:
\begin{equation*}
  {\tilde I}_N^{MC}=\frac{b-a}{N}\sum_{j=1}^N f(x_j).
\end{equation*}
For integration over probability distributions, we assign $\mu_j$ to
its expectation value $\mu_j=1/N$ and the MC integral becomes
\begin{equation}
  {\tilde I}_N^{MC}=\frac{1}{N}\sum_{j=1}^N f(x_j),
  \label{eq:mci}
\end{equation}
although intuition suggests that the explicit form from equation
(\ref{eq:ItildeN}) will yield a better result. 

The development leading to equation (\ref{eq:ItildeN}) remains nearly
unchanged for a multidimensional integral:
\begin{equation}
  I = \int_{\mathbb{R}^k} f(\bx) g(\bx) d^kx.
\end{equation}
As in the one-dimensional case, the Lebesgue integral becomes
\begin{eqnarray}
  I &=& \int_{\mathbb{R}^k} f(\bx)g(\bx)\,d^kx \nonumber \\
  &=& M(y_0) y_0 + \lim_{N\rightarrow\infty}{\cal L}_{{\cal S}} 
  = M(y_0) y_0 + \lim_{N\rightarrow\infty}{\cal U}_{{\cal S}}
  \nonumber \\
  &=& M(y_0) y_0 + \int_{y_0}^{y_N} M(y)\,dy.
\label{eq:lebesgue3}
\end{eqnarray}
where the only difference is that $M(y)$ is now the measure of the set
of points with $y\le f(\bx)$ with $\bx\in\mathbb{R}^k$.  Perhaps more
remarkable than the similarity of equation (\ref{eq:lebesgue2}) with
equation (\ref{eq:lebesgue3}) is that the numerical Lebesgue integral
is one-dimensional independent of the dimensionality $k$.  However,
this does not simplify the computational work; one still needs to
sample a subset of $\mathbb{R}^k$ to evaluate equation
(\ref{eq:lebesgue3}).  The MC version proceeds similarly as well:
replace $x_j$ by ${\bf x}_j$ in equation (\ref{eq:mci}).

In summary, Monte Carlo integration is most often posed as the
expectation over a distribution, which, more generally, is a Lebesgue
integral.  Lebesgue integration and differentiation theory suggests
alternative computational approaches akin to traditional Riemann-based
numerical analysis, if the underlying likelihood function and prior
probability density are well-behaved functions.  We will see in the
next section that a truncation-error criterion applied to the marginal
likelihood integral in the form of equation (\ref{eq:lebesgue3}) can
improve the HMA.

\section{Application to the marginal likelihood integral}
\label{sec:evid}

Now, given a MC-computed sample from the posterior distribution
$P(\param|\data)$ with prior distribution $\pi(\param)$ and likelihood
function $L(\data|\param)$, how does one compute the marginal
likelihood?  The integral in equation (\ref{eq:evidence}) states that
marginal likelihood is the expectation of the likelihood with respect
to the prior distribution.  This is the same as equation
(\ref{eq:lebesgue3}) with $\param\in\Omega_s\subset\mathbb{R}^k$
replacing $\bx$, $L(\data|\param)$ replacing $f(\bx)$, $\pi(\param)$
replacing $g(\bx)$. Alternatively, returning to equation
(\ref{eq:Zdef0}), the integral $Z\equiv P(\data)$ is implicitly
defined by
\begin{equation}
  P(\data) \int_{\Omega_s}
  \frac{d\param\,P(\param|\data)}{L(\data|\param)} = \int_{\Omega_s}
  d\param\,\pi(\param) \equiv J.
  \label{eq:Zdef}
\end{equation}
The value $J$ is the probability of $\pi$ over $\Omega_s$.  We will
assume that $\Omega_s\subseteq\Omega$; this implies that $J\le1$
since $\int_\Omega d\param\pi(\param) = 1$.  In addition, the
existence of equation (\ref{eq:Zdef}) implies that $L(\data|\param)>0$
almost everywhere in $\Omega$.  Defining $Y \equiv L^{-1}$, it follows
that the Lebesgue integral of the integral on the left-hand-side of
equation (\ref{eq:Zdef}) is
\begin{equation}
  K \equiv \int_{\Omega_s}
  \frac{d\param\,P(\param|\data)}{L(\data|\param)} = \int
  M(Y)\,dY + M(Y_0)Y_0
  \label{eq:evidenceX}
\end{equation}
with measure
\begin{equation}
  M(y) = \int_{Y(\data|\param)>y}d\param\, P(\param|\data).
    \label{eq:measureX}
\end{equation}
Intuitively, one may interpret this construction as follows: divide up
the parameter space $\param\in\Omega_s\subset\mathbb{R}^k$ into volume
elements sufficiently small that $P(\param|\data)$ is approximately
constant.  Then, sort these volume elements by their value of
$Y(\data|\param) = L^{-1}(\data|\param)$.  The probability element
$dM\equiv M(Y+dY) - M(Y)$ is the prior probability of the volume
between $Y$ and $Y + dY$.

Clearly $M(Y)\in[0,1]$ and may be trivially computed from a
MCMC-generated posterior distribution.  Using our finite MC-sampled
distributed as the posterior probability, $\param\sim
P(\param|\data)$, and converting the integral to a sum, we have the
following simple estimate for $M_i\equiv M(Y_i)$:
\begin{equation}
  M_i^{[l]} \equiv \frac1N \sum_{j=1}^N \mathbf{1}_{\{Y_j>Y_i\}},
  \qquad
  M_i^{[u]} \equiv \frac1N \sum_{j=1}^N \mathbf{1}_{\{Y_j\ge Y_i\}},
  \qquad
  M_i \equiv \frac{M_i^{[l]} + M_i^{[u]}}{2},
  \label{eq:measureX3}
\end{equation}
where we have defined the left and right end points from equation
(\ref{eq:nummeas2}) and the mean separately so that $M_i^{[l]} \le M_i
\le M_i^{[u]}$.  The indicator function $\mathbf{1}_{\{\}}$ enforces
the inclusion of a contribution $1/N$ for index $j$ only if
$\{Y_j>Y_i\}$ or $\{Y_j\ge Y_i\}$ for the lower and upper form,
respectively.  Alternatively, these sums may be expressed using
equations (\ref{eq:nummeas})--(\ref{eq:ItildeN}).  

We may now estimate the marginal likelihood from equation
(\ref{eq:Zdef}) using the second part of equation (\ref{eq:lebesgue3})
for finite $N$ by gathering terms in $Y_i$ to get
\begin{eqnarray}
  K &\equiv& \int M(Y)\,dY + M(Y_0)Y_0 \label{eq:intK} \\ 
  &\approx& {\tilde K} \equiv
  \sum_{i=0}^N\left(Y_{i+1} - Y_{i}\right)M_{i} + \frac{M_0}{L_0}
  =
  \sum_{i=0}^N \left(\frac{1}{L_{i+1}} - \frac{1}{L_{i}}\right)M_{i} +
  \frac{M_0}{L_0} \label{eq:numericK} \\ &=& \sum_{i=1}^N \frac{1}{L_{i}}M_{i-1}
  - \sum_{i=1}^N \frac{1}{L_{i}}M_{i} = \sum_{i=1}^N
  \frac{1}{L_{i}}(M_{i-1} - M_i) = \frac1N \sum_{i=1}^N \frac1{L_{i}}.
  \label{eq:harmonic0}
\end{eqnarray}
In deriving equation (\ref{eq:harmonic0}), the leading term
$M(Y_0)Y_0$ from equation (\ref{eq:lebesgue3}) is absorbed into the
sum and $M(Y_N)=0$.  Assuming that $J=1$ in equation (\ref{eq:Zdef})
and using equation (\ref{eq:harmonic0}) yields
\begin{equation}
  {\tilde Z} \equiv {\tilde P}(\data) = {\tilde J}/{\tilde K} =
  \left({\displaystyle\frac1N \sum_j \frac{1}{L_j}}\right)^{-1}.
  \label{eq:harmonic}
\end{equation}
This is an alternative derivation for the ``harmonic mean''
approximation (HMA) to the marginal likelihood.

\subsection{Convergence of $K$}
\label{sec:gauss_ex}

The evaluation of the integral $K$ (eq. \ref{eq:intK}) may fail both
due to insufficient sampling and intrinsic divergence.  As an example
of the former, a sparse sampling may lead to large intervals
$Y_{i+1}-Y_{i}$ and inflated truncation error (eq. \ref{eq:numericK}).
We will consider this further in the next section.  As an example of
the latter, consider the textbook inference of an unknown mean
$\theta$ from a sample of $N$ normally distributed points
$x\sim{\cal N}(\theta, \sigma_x^2)$.  The likelihood function is
\begin{equation}
  L(D|\theta) =
  \prod_{i=1}^N\frac{e^{-(x_i-\theta)^2/2\sigma_x^2}}{\sqrt{2\pi\sigma_x^2}} 
  = L_0 e^{-({\bar x} - \theta)^2N/2\sigma_x^2}
  \label{eq:likeex}
\end{equation}
where $L_0=\sup\{L: x\in\mathbb{R}\}$ and ${\bar x}$ is the sample
mean.  Let the prior distribution for $\theta$ be ${\cal N}(\theta_0,
\sigma_\theta^2)$. We use an MCMC algorithm to sample the posterior
distribution $\theta$.

Now, let us evaluate $K$ using Lebesgue integration for this example.
To begin, we need the measure function $M(Y)=\int_{L^{-1}(D|\theta)>Y}
dP$ with $Y=L^{-1}$ We may use equation (\ref{eq:likeex}) to solve for
$\theta = \theta(L)$, noting that the solution has two branches.
After some algebra, we have:
\begin{equation}
  M(Y) = 1 - \frac12\left[
    \mbox{erf}\left(\frac{ {\bar x} - {\bar y} + \Delta(Y)
      }{\sqrt{2{\bar\sigma^2}}}\right)
    - 
    \mbox{erf}\left(\frac{ {\bar x} - {\bar y} - \Delta(Y)
      }{\sqrt{2{\bar\sigma^2}}}\right) 
  \right]
  \label{eq:Myg}
\end{equation}
where
\[ {\bar y} \equiv \frac{\sigma_x^2\theta_0/N +
  \sigma_\theta^2 {\bar x}}{\sigma_x^2/N + \sigma_\theta^2},
\qquad {\bar \sigma^2} \equiv
\frac{\sigma_\theta^2\sigma_x^2/N}{\sigma_x^2/N +
  \sigma_\theta^2} = \left(\frac{1}{\sigma_\theta^2} +
  \frac{N}{\sigma_x^2}\right)^{-1},
\]
\[
Y_0 \equiv L_0^{-1}, \qquad \Delta(Y) \equiv
\sqrt{\frac{2\sigma_x^2}{N}\log(Y/Y_0)}.
\]
The value ${\bar y}$ is the variance weighted mean of the prior mean
and the sample mean, and ${\bar\sigma^2}$ is the harmonic mean of the
variance of the prior distribution and the variance of the sample
mean.  The value $Y_0$ is the minimum value for $Y$ and
$\Delta(\cdot)$ describes the offset of $\theta$ with increasing $Y$.
Note that $\Delta(Y_0)=0$.

\begin{figure}[thb]
\centering
\includegraphics[width=0.6\textwidth]{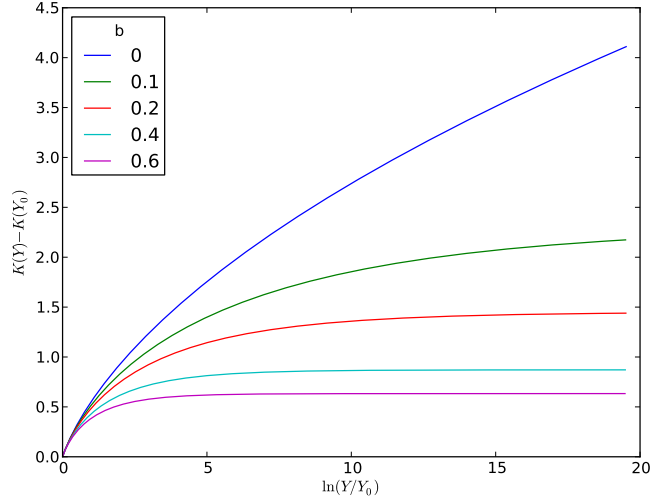}
\caption{\label{fig:gauss_ex1} The integral $K(Y)-K(Y_0)$ is shown as
  a function of $Y/Y_0$ for various values of the ratio
  $b\equiv\sigma_x^2/(N\sigma_\theta^2)$.  For an uninformative prior
  distribution $\sigma_\theta\gg \sigma_x/N$ and $b\rightarrow0$ and
  $K(Y)$ diverges with increasing $Y/Y_0$.  For an informative prior
  distribution ${\cal O}(b)\sim 1$ and $K(Y)$ convergences quickly
  with $Y/Y_0$.}
\end{figure}

Since the values of $Y$ are obtained by a sampling, the sample will
not cover $[Y_0, \infty)$ but will be limited from above by the
smallest sampled value of the likelihood $Y_{max}=L_{min}^{-1}$.  We
define this limited value of the Integral $K$ as
\begin{equation}
  K(Y_{max}=L_{min}^{-1}) \equiv \int_{Y_0}^{Y_{max}} dY M(Y) +
  M(Y_0)Y_0 = \int_{Y_0}^{Y_{max}} dY M(Y) + Y_0
  \label{eq:Kyg}
\end{equation}
where the last equality uses $M(Y_0)\equiv1$.  Clearly $K = K(\infty)
> K(Y_{max})$.  The magnitude of the truncation due to finite
sampling, $K(\infty) - K(Y_{max})$, depends critically on the width of
the likelihood distribution relative to the prior distribution.  We
describe this ratio of widths by
$b\equiv\sigma_x^2/(N\sigma_\theta^2)$.  The convergence condition
$\lim_{Y\rightarrow\infty}[K(\infty)-K(Y_{max})]\rightarrow0$ requires
that $M(Y)$ decreases faster than $Y^{-1-\epsilon}$ for some
$\epsilon>0$.  For $b=0$ and large $Y$, $\int^Y dY M(Y)$ increases as
$\log(\log Y)$.  For $b>0$, $\int^Y dY M(Y)$ decreases at least as
fast $Y^{-b}$.  Figure \ref{fig:gauss_ex1} shows $K(\infty) - K(Y_0)$
as a function of $b$ and suggests that $b>0.1$ is sufficient to obtain
convergence for practical values of $N$.  Qualitatively, a prior
distribution that limits $Y$ from above (or, equivalently, $L$ from
below) will prevent the divergence.

\begin{figure}[thbp]
\centering
\subfigure[$b=0.2$]{\includegraphics[width=0.6\textwidth]{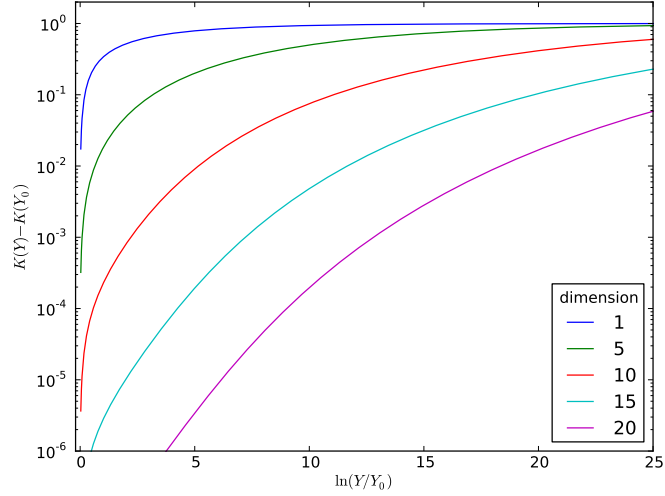}}
\subfigure[$b=0.6$]{\includegraphics[width=0.6\textwidth]{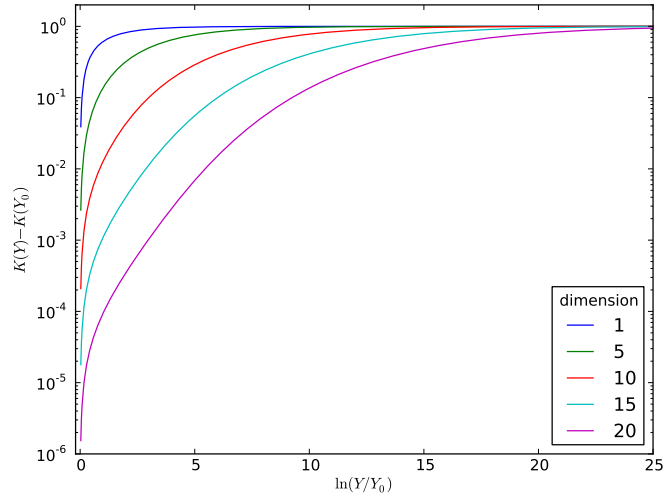}}
\caption{\label{fig:gauss_ex2} As in Figure
  \protect{\ref{fig:gauss_ex1}}, the integral $K(Y)-K(Y_0)$ is shown
  as a function of $Y/Y_0$ for the ratio $b=0.2, 0.6$ for
  $k$-dimensional normal data distributions.  The integral $K(Y)$
  converges more rapidly with increasing $b$ (as in
  Fig. \protect{\ref{fig:gauss_ex1}}) but increasingly slowly with
  $k$.  The run of $K(Y)-K(Y_0)$ is normalized to 1 at $Y=\infty$ to
  facilitate comparison.}
\end{figure}

Similar asymptotic expressions for $K(Y)$ may be derived for
multivariate normal distributions.  For simplicity, assume that data
is distributed identically in each of $k$ dimensions and, for ease of
evaluation, take $\mathbf{\bar x} = \mathbf{\bar y}$.  Then
\begin{equation}
  M(Y) = 
  \Gamma\left(\frac{k}{2}, (1 +
    b)\log(Y/Y_0)\right)/\Gamma\left(\frac{k}{2}\right)
  \label{eq:muliMyg}
\end{equation}
where $\Gamma(a, x)$ is the upper incomplete gamma function and
$\Gamma(a)$ is the complete gamma function.  Using the standard
asymptotic formula for $\Gamma(a, x)$ in the limit of large $Y$, one
finds that
\begin{equation}
  M(Y) \rightarrow \frac{1}{\Gamma(k/2)}
  \left[\sqrt{1+b^2}\log\left(\frac{Y}{Y_0}\right)\right]^{k/2 - 1}
  \left(\frac{Y}{Y_0}\right)^{-1-b} \qquad \mbox{for\ } Y\gg k/2.  
  \label{eq:muliMygasymp}
\end{equation}
This expression reduces to equation (\ref{eq:Myg}) when $k=1$, but
more importantly, this shows that the tail magnitude of $M(Y)$
increases with dimension $k$.  Figure \ref{fig:gauss_ex2} illustrates
this for various values of $k$ and $b$.

The divergence of $K$ in the limit $\sigma_\theta^2\rightarrow\infty$
shows that normally distributed likelihood function with an
uninformative prior is divergent.  Moreover, Figures
\ref{fig:gauss_ex1} and \ref{fig:gauss_ex2} further demonstrate that
weakly informative prior with very small $b$ is likely to be
numerically divergent even if $K$ formally converges.  Intuitively,
the cause is clear: if the Markov chain never samples the wings of the
prior distribution that still make significant contribution to $K$,
then $K$ will increase with sample size.  Analytically, the failure is
caused by the measure decreasing too slowly as $Y$ increases (as
described at the beginning of \S\ref{sec:evid}).  Empirically, the
convergence of the integral $K$ may be tested by examining the run of
$K(Y)$ for increasing $Y$.

\subsection{Application to the MCMC posterior distribution}
\label{sec:newevid}

Overall, \S\ref{sec:evid} highlights the marginal likelihood as a
numerical quadrature.  We have considered the path of approximations
leading from the quadrature to standard expression for the HMA.  We
have also considered the intrinsic requirements on the prior
distribution so that the meausure $M(Y)$ is convergent.  This
development suggests that there are two sources of error in evaluating
equation (\ref{eq:Zdef}) using equations (\ref{eq:evidenceX}) and
(\ref{eq:numericK}).  The first is a truncation error of order
$(Y_{i+1} - Y_{i})^2$.  The second is a bias error resulting from
specifying the computational grid by a Monte Carlo procedure.  We will
consider each in turn.

Thinking again in terms of numerical quadrature, we can trust the sum
in equation (\ref{eq:numericK}) when adjacent values of $L_j$ are
close.  The error in ${\tilde K}$ will be dominated by the extremely
small values of $L_i$ that lead to $h_i\equiv Y_{i+1} -Y_{i}\gg1$.
Specifically, the error for such a term will be $\propto h_i^2$, and a
small number of such terms can dominate the error for ${\tilde K}$.
Numerical analysis is based on functional approximation of smooth,
continuous, and differentiable functions.  This truncation error
estimate assumes that $M(Y)$ is such a function.  Although not true
everywhere, we have argued in \S\ref{sec:intgr}) that this assumption
should be valid over a countable number of intervals in practice.  In
addition, we expect the sample to be strongly clustered about the
value of the posterior mode.  By the Central Limit Theorem for a
likelihood dominated posterior, the distribution of $\theta$ tends
toward a normal distribution.  Therefore, larger $N$ will yield more
extremal values of $L$ and \emph{increase} the divergence of
$h_i=L^{-1}_{i+1} - L^{-1}_{i}$.  Eventually, for proper prior
distributions and sufficiently large $N$, the smallest possible value
of $L$ will be realized as long as $L>0$ (see eq. \ref{eq:Zdef} and
following discussion).  Further sampling will reduce the largest
intervals, and this will lead to the decrease of large $h_i$, and
finally, convergence.

The second source of error is closely related to the first.  After
eliminating the divergent samples with $h_i\gg1$, the sampled domain
$\Omega_s$ will be a subset of the originally defined domain
$\Omega_s\subset\Omega$.  That is, the MCMC sample will not cover all
possible values of the parameter vector $\mathbf{\theta}$. This
implies that the numerical quadrature of equation (\ref{eq:Zdef}) will
yield ${\tilde J}<1$.  Identification of these error sources
immediately suggests solutions.  Note that this observation does not
change the problem definition in some new way, but rather, allows us
to exploit the MCMC-chosen domain $\Omega_s$ to eliminate the
divergence for small $b$ described in \S\ref{sec:gauss_ex}.

First, we may decrease the truncation error in ${\tilde K}$ by
ordering the posterior sample by increasing values of $Y$ and
truncating the sequence at the point where $h_i>h_\ast$ for some
choice $h_\ast\ll1$.  Next, we need a consistent evaluation for
${\tilde J}$.  We may use the sampled posterior distribution itself to
estimate the sampled volume in $\Omega_s\subset\Omega$.  This may be
done straightforwardly using a space partitioning structure.  A
computationally efficient structure is a binary space partition (BSP)
tree, which divides a region of parameter space into two subregions at
each node.  The most easily implemented tree of this type for
arbitrary dimension is the kd-tree (short for k-dimensional tree).
The computational complexity for building the tree from the $N$
sampled points in parameter space scales as ${\cal O}(N\log^2N)$ using
the Quicksort algorithm at each successive level \citep[this may be
improved, see][]{Cormen.etal:01}.  Each leaf has zero volume.  Each
non-leaf node has the minimum volume enclosing the points in the node
by coordinate planes.  Assigning the volume containing a fixed number
of leaves $\barm$ (e.g. ${\bar m}=16$ or $32$), and some
representative value of the prior probability in each node (such as a
$p$-quantile or mean value), one may immediately sum product of each
volume and value to yield an estimate of $J$.  For modest values $N$,
we will almost certainly find that ${\tilde J}<1$.  Since the MCMC
chain provides the values of $\pi(\param)$ and
$P(\theta)=\pi(\param)L(\data|\param)$, we may use the same tree to
evaluate both ${\tilde Z}$ and ${\tilde J}$ over the sampled volume
$\Omega_s$.

The example in \S\ref{sec:gauss_ex} suggests that evaluation of $K$
may stymied by poor convergence unless the prior distribution is
restrictive. Therefore, if the value of $K$ is divergent or very
slowly convergent, the evaluation of $Z$ using $K/J$ will fail whether
or not we use the improved truncation criteria.  Direct evaluation of
the $Z$ is free from this divergence and remains an practical option
in this case.  The advantage of a direct evaluation is clear: the
converged Markov chain samples the domain $\Omega$ proportional to the
integrand of equation (\ref{eq:evidence}), and therefore, we expect
\[
\lim_{N\rightarrow\infty}\int_{\Omega_s}d\param\pi(\param)L(\data|\param)\gg
\lim_{N\rightarrow\infty}\int_{\Omega\setminus\Omega_s}d\param\pi(\param)L(\data|\param)\rightarrow0
\]
for large sample size by construction.  We propose a hybrid of
cubature and Monte Carlo integration.  The BSP tree provides a
\emph{tiling} of multidimensional volume by using the posterior
distribution to define volume elements, $\Delta V$.  We use a
$p$-quantile (such as the $p=0.5$ median) or mean value of the
posterior probability or the prior probability to assign a probability
value to each volume element.  An approximation to the integrals $Z$
and $J$ follow from summing the field values over the volume elements,
analogous to a multidimensional Riemann rule.

Although $\pi(\param)L(\data|\param)\Delta V = \mbox{constant}$ for a
\emph{infinite} posterior sample, there are several sources of error
in practice.  First, the variance in the tessellated parameter-space
volume will increase with increasing volume and decreasing posterior
probability.  This variance may be estimated by bootstrap.  Secondly,
the truncation error of the cubature increases with the number of
points per element.  As usual, there is a variance--bias trade off
choosing the resolution of the tiling: the bias of the probability
value estimate increases and the variance decreases as the number of
sample points per volume element increases.  The prior probability
value will be slowing varying over the posterior sample for a typical
likelihood-dominated posterior distribution, so the bias will be
small.  This suggests that larger numbers of points per cell will be
better for the evaluation of $J$ and a smaller number will be better
for $Z$.  Some practical examples suggest that the resulting estimates
are not strongly sensitive to the number of points per cell (${\bar
  m}=16$ or 32 appears to be a good compromise).  Almost certainly,
there will be a bias toward larger volume and therefore larger values
of ${\tilde Z}$ and this bias will increase with dimension most
likely.

To summarize, we have described two approaches for numerically
computing $Z$ from a MCMC posterior simulation.  The first evaluates
of the integral $K$ by numerical Lebesgue integration, and the second
evaluates $Z$ directly by a parameter space partition obtained from
the sampled posterior distribution.  The first is closely related to
the HMA.  It applies ideas of numerical analysis the integral that
defines the HMA.  The second is more closely related to the Laplace
approximation.  In some sense, Laplace approximation is an integral of
a parametric fit to the posterior distribution.  The tree integration
described above is, in essence, an integral of a non-parametric fit to
the posterior distribution.  The advantage of the first method its
amenability to analysis.  The disadvantage is the limitation on
convergence as illustrated in \S\ref{sec:gauss_ex}.  The advantage of
the second method is its guaranteed convergence.  The disadvantage is
its clear, intrinsic bias and variance.  The variance could be
decreased, presumably, using high-dimensional Voronoi triangulation
but not without dramatic computational cost.

\subsection{Discussion of previous work on the HMA}

The HMA is treated as an expectation value in the literature.  One of
the failures pointed out by \citet{Meng.Wong:96} and others is that
the HMA is particularly awful when the sample is a single datum.  In
the context of the numerical arguments here, this is no surprise: one
cannot accurately evaluate a quadrature with a single point!  Even for
larger samples, the HMA is formally unstable in the limit of a
thin-tailed likelihood function owing to the divergence of the
variance of the HMA.  \citet{Raftery.etal:07} address this failure of
the statistic directly, proposing methods for stabilizing the harmonic
mean estimator by reducing the parameter space to yield heavier-tailed
densities.  This is a good alternative to the analysis presented here
when such stabilization is feasible.  As previously mentioned,
\citet{Wolpert:02} presents conditions on the posterior distribution
for the consistency of the HMA.  Intuitively, there is a duality with
the current approach.  Trimming the Lebesgue quadrature sum so that
the interval $Y_{i+1} - Y_{i} < h_\ast$ is equivalent to lopping off
the poorly sampled tail of the posterior distribution.  This
truncation will be one-sided in the estimate of the marginal
likelihood ${\tilde Z}$ since it removes some of the sample space.
However, this may be compensated by an appropriate estimation of
${\tilde J}$.

\section{The new algorithms}
\label{sec:algo}

The exposition and development in \S\ref{sec:evid} identifies the
culprits in the failure of the HMA: (1) truncation error in the
evaluation of the measure $M(Y)$; and (2) the erroneous assumption
that $J=1$ when $\Omega_s\subset\Omega$ in practice. We now present
two new algorithms, the \emph{Numerical Lebesgue Algorithm} (NLA) and
the \emph{Volume Tessellation Algorithm} (VTA), that implement the
strategies described in \S\ref{sec:newevid} to diagnose and mitigate
this error.  NLA computes ${\tilde K}$ and VTA computes ${\tilde J}$
and, optionally, ${\tilde Z}$ directly from equation
(\ref{eq:evidence}).  In the following sections, we assume that
$\Omega\subset\mathbb{R}^k$.

\subsection{Description}

We begin with a converged MCMC sample from the posterior distribution.
After the initial sort in the values of $L_j$, the NLA computes the
difference $h_j=L^{-1}_{j-1} - L^{-1}_{j}$ with $j=N, N-1, \ldots$ to
find the first value of $j=n$ satisfying $h_j<h_\ast$.  The
algorithm then computes the $M_i$ for $i=n,\ldots, N$ using equation
(\ref{eq:measureX3}).  For completeness, we compute the $M_i$ using
both the restriction $L(\data|\param)>L_i$ and $L(\data|\param)\ge
L_i$ to obtain lower and upper estimate for $M(Y)$.  Then, these may
be combined with ${\cal L}_S$ and ${\cal U}_S$ from \S\ref{sec:intgr}
to Riemann-like upper and lower bounds on ${\tilde K}$ .  See the
listing below for details.  Excepting the sort, the work required to
implement this algorithm is only slightly harder than the HMA.

The VTA uses a kd-tree to partition the $N$ samples from posterior
distribution into a spatial regions.  These tree algorithms split
$\mathbb{R}^k$ on planes perpendicular to one of the coordinate system
axes.  The implementation described here uses the median value along
one of axes (a \emph{balanced} kd-tree).  This differs from general
BSP trees, in which arbitrary splitting planes can be used.  There
are, no doubt, better choices for space partitioning such as Voronoi
tessellation as previously discussed, but the kd-tree is fast, easy to
implement, and published libraries for arbitrary dimensionality are
available.  Traditionally, every node of a kd-tree, from the root to
the leaves, stores a point.  In the implementation used here, the
points are stored in leaf nodes only, although each splitting plane
still goes through one of the points.  This choice facilitates the
computation of the volume spanned by the points for each node as
follows.  Let $\barm_j$ be the number of parameter-space points
$\param^{[n]}, n=1,\ldots,{\bar m}_j$ in the $j^{\mbox{th}}$ node.
Let ${\bf f}^{[n]}$ denote the field quantities at each point
$\param^{[n]}$.  Some relevant field quantities include the values of
the unnormalized posterior probability and the prior probability.  The
volume for Node $j$ is
\begin{equation}
  V_j = \prod_{i=1}^k
  \left[\max(\theta^{[1]}_i,\ldots,\theta^{[\barm_j]}_i) -
    \min(\theta^{[1]}_i,\ldots,\theta^{[\barm_j]}_i) \right].
\label{eq:volj}
\end{equation}
The set of nodes with $\barm=\barm_j=2^q$ for some fixed integer $q$,
determines an exclusive volume partition of the parameter space
spanned by the point set, the \emph{frontier}.  The value of $q$ is
chosen large enough to limit the sampling bias of field quantities in
the volume but small enough resolve the posterior modes of interest.
The values $q\in[2,\ldots,6]$ seem to be good choices for many
applications.  Each node in the frontier is assigned a representative
value ${\bf f}_\ast$.  I use $p$-quantiles with $p=0.1, 0.5, 0.9$ for
tests here.  The resulting estimate of the integrals ${\tilde J}$
and/or ${\tilde Z}$ follow from summing the product of the frontier
volumes with their values ${\bf f}_\ast$.

\subsection{Performance}

Both the NLA and the VTA begin with a sort of the likelihood sequence
$L$ and this scales as ${\cal O}(N\log N)$.  In the NLA, the
computation of the $M_k$ followed by the computation of ${\tilde Z}$
is ${\cal O}(N)$.  The sequence $\{M_k\}$ is useful also for diagnosis
as we will explore in the next section.  However, in many cases, we do
not need the individual $M_i$ but only need the differential value
$M_i - M_{i+1}$ to compute ${\tilde Z}$, which contains a single term.
The values of likelihood may range over many orders of magnitude.
Owing to the finite mantissa, the differential value be necessary to
achieve adequate precision for large $N$, and the NLA may be modified
accordingly.  The algorithm computes the lower, upper, and
trapezoid-rule sums (eqns. \ref{eq:LUn}--\ref{eq:Tn}) for the final
integral ${\tilde Z}$.  For large posterior samples, e.g. $N>10000$,
the differences between ${\cal L}_{{\cal S}}$ and ${\cal U}_{{\cal
    S}}$ are small.  Indeed, a more useful error estimate may be
obtained by a random partitioning and subsampling of the original
sequence $\{L_k\}$ to estimate the distribution of ${\tilde Z}$ (see
the examples in \S\ref{sec:examples}).  In practice, computing the
marginal likelihood from a posterior sample with $N=400000$ takes 0.2
CPU seconds on a single 2Ghz Opteron processor.  Although NLA could be
easily parallelized over $n$ processors to reduce the total runtime by
$1/n$ this seems unnecessary.

The kd-tree construction in VTA scales as ${\cal O}(kN\log^2 N)$
followed by a tree walk to sum over differential node volumes to
obtain the final integral estimates that scales as ${\cal O}(N\log
N)$.  This scaling was confirmed empirically using the
multidimensional example described in \S\ref{sec:highD} with dimension
$k\in[1,40]$ and sample size $N\in[1000,10000000]$.  Computing the
marginal likelihood from a posterior sample with $N=400000$ and $k=10$
takes 4.4 CPU seconds on a single 2Ghz Opteron processor, and,
therefore, the computation is unlikely to be an analysis bottleneck,
even when resampling to produce a variance estimate.  The leading
coefficient appears to vary quite weakly the distribution, although
there may be undiscovered pathological cases that significantly
degrade performance.  The required value of $N$ increases with
parameter dimension $k$; $N=400000$ is barely sufficient for $k=40$ in
tests below.  Subsampling recommends the use of even larger chains to
mitigate dependence of the samples.  Therefore, the first practical
hardware limitation is likely to be sufficient RAM to keep the data in
core.

\renewcommand{\algorithmiccomment}[1]{ // #1}
\renewcommand{\thealgorithm}{NL.}

\begin{algorithm*}
  \caption{Algorithm to compute the marginal likelihood from a
    posterior sample by Lebesgue integration.  In Line 3, the value of
    $h_\ast$ can be chosen to trim anomalous values of likelihood.
    In Line 9, both choices for the inequality in $\Theta_j$ (see
    eq. \ref{eq:nummeas2}) can be computed and stored simultaneously.
    Combined with Line 18, we can bracket the values by lower and
    upper sums (algorithmic error).}
  \label{alg:1}
  \begin{algorithmic}[1]
    \REQUIRE Likelihood values $\{L_j\}, j=1,\ldots,N$ from the
    simulated posterior distribution
    \STATE Sort the sequence so that $\{L_j\ge L_{j-1}\}$
    \STATE $n=N$    \COMMENT{find the smallest $n$ obeying the
      threshold condition}
    \WHILE{$L_{n-1}^{-1} - L^{-1}_{n} < h_\ast$}
    \STATE $n\leftarrow n-1$
    \ENDWHILE
    \FOR{$i=n$ to $N$}
    \STATE $M_i\leftarrow 0$
    \COMMENT{compute the measure, see eq. \ref{eq:measureX3}}
    \FOR{$j=i$ to $N$}
    \STATE $M_i \leftarrow M_i + 1/L_j$
    and/or $M_i \leftarrow M_i + 1/L_{j+1}$
    \ENDFOR
    \STATE Save the values $M_i$
    \ENDFOR
    \FOR{$i=n$ to $N$} 
    \STATE $M_i\leftarrow M_i/M_N$
    \COMMENT{normalize the prior measure}
    \ENDFOR
    \STATE ${\tilde Z}\leftarrow 0$
    \COMMENT{compute the marginal likelihood, see
      eqns. \ref{eq:LUn}--\ref{eq:Tn}}
    \FOR{$i=n$ to $N$}
    \STATE $
           {\tilde Z} \leftarrow {\tilde Z}
           +
           \left(M_i - M_{i+1}\right)
           \left\{\begin{array}{ll}
           L_i & \mbox{lower sum} \\
           L_{i+1} & \mbox{upper sum} \\
           (L_i + L_{i+1})/2 & \mbox{trapezoidal rule}\\
           \end{array}\right\}
           $
    \ENDFOR
    \STATE Save the estimated marginal likelihood, ${\tilde Z}$ and
    algorithmic error
  \end{algorithmic}
\end{algorithm*}

\renewcommand{\thealgorithm}{VT.}

\begin{algorithm}
  \caption{Algorithm to estimate $J=\int_{\Omega_s}
    d\param\,\pi(\param|\model)$ over the domain
    $\Omega_s\subset\Omega$ sampled by the MCMC algorithm using the
    space partitioning kd-tree.}
  \label{alg:2}
  \begin{algorithmic}[1]
    \REQUIRE Likelihood values $\{L_j\}, j=1,\ldots,N$ from the
    simulated posterior distribution
    \STATE Change variables $Y_j = 1/L_j$
    \STATE Sort the sequence so that $\{Y_{j-1}\le Y_{j}\}$
    \STATE Create an empty point set ${\cal P}$
    \STATE $n\leftarrow1$  \COMMENT{find the largest $n$ obeying the
      threshold condition}
    \WHILE{$Y_{n+1} - Y_{n} < h_\ast$}
    \STATE $n\leftarrow n+1$
    \STATE Add $(\param^{[n]}, {\bf f}^{[n]})$ to ${\cal P}$
    \ENDWHILE
    \STATE $\mbox{Node}_{root} = \mbox{BuildKD}({\cal P})$
    \STATE Find the set of frontier nodes ${\cal F}$ with the desired
    number of points $\barm$
    \STATE ${\tilde G} \leftarrow 0$
    \FOR{each node $j\in{\cal F}$}
    \STATE Compute the median value of ${\bf f}^{[n]}$ among the $\barm$
    points
    \STATE ${\tilde G}\leftarrow{\tilde G} + V_j\times\mbox{median}({\bf f})$
    \ENDFOR
    \STATE Save the estimated value of the integrals ${\tilde G}$
  \end{algorithmic}

  \bigskip
  {\bf Procedure:} BuildKD(${\cal P}$)\newline
  \vspace{-16pt}
  \begin{algorithmic}[1]
    \REQUIRE A set posterior points and values 
    $(\param^{[n]}, {\bf f}^{[n]})\in{\cal P}$
    \IF{${\cal P}$ contains only one member}
    \RETURN{pointer to this leaf}
    \ELSE
    \STATE Compute and store the range for each coordinate and the
    volume for this node
    \STATE Locate the coordinate dimension $j\in1\ldots k$ with 
    maximum variance
    \STATE Determine the median value of $\theta_j{[i]}$ (e.g. by Quicksort)
    \STATE Split ${\cal P}$ into two subsets by the hyperplane
    defined by $\mbox{median}(\theta_k)$: ${\cal P}_{left}, {\cal P}_{right}$.
    \STATE $\mbox{Node}_{left}\  = \mbox{BuildKD}({\cal P}_{left})$
    \STATE $\mbox{Node}_{right} = \mbox{BuildKD}({\cal P}_{right})$
    \ENDIF
    \RETURN{pointer to the root node of the constructed tree}
  \end{algorithmic}
\end{algorithm}

\section{Tests \& Examples}
\label{sec:examples}

To estimate the marginal likelihood using the methods of the previous
section, we may use either the NLA to estimate $K$ and the VTA to
estimate $J$ or use the VTA alone to estimate $Z$.  Examples below
explore the performance of these strategies.  The MCMC posterior
simulations are all computed using the UMass Bayesian Inference Engine
\citep[BIE,][]{Weinberg:10}, a general-purpose parallel software
platform for Bayesian computation.  All examples except that in
\S\ref{sec:highD} simulate the posterior distribution using the
parallel tempering scheme \citep{Geyer:91} with $T=128$ and 20
temperature levels.  Convergence was assessed using the subsampling
algorithm described in \citep{GVDP:99}, a generalization of the
\citet{Gelman.Rubin:92} test.

\subsection{Fidelity of the NLA and the VTA}
\label{sec:test}

For a simple initial example, let us compute the marginal likelihood
for a data sample $\data$ of 100 points $x \sim {\cal N}(0.5, 0.03)$
modelled by ${\cal N}(\theta, 0.03)$ with prior distribution for
$\theta \sim {\cal U}(-0.2, 1.2)$.  The marginal likelihood $Z$ can be
computed analytically from $\data$ for this simple example. The final
200,000 converged states of the MCMC-generated chain were retained.
Application of the NLA for ${\tilde K}$ and the VTA for ${\tilde J}$
gives a value of $\log{\tilde Z}=31.15\pm0.02$ (95\% confidence
interval), close to but systematically smaller than the analytic
result: $\log Z=31.36$.  A value of $h_\ast=0.05$ seems appropriate
from numerical considerations, although experiments suggest that the
algorithm is not sensitive to this choice as long as $h_\ast$ is not
so small to decimate the sample or so large that error-prone outliers
are included.  It is prudent to check a range of $h_\ast$ to determine
the appropriate value of each problem.  The VTA yields $\log{\tilde
  Z}=31.34\pm0.01$, consistent with the analytic result.  The bias in
the first estimate appears to be caused by an overestimate of ${\tilde
  J}$ produced by the VTA.  This might be improved by a space
partition whose cells have smaller surface to volumes ratios
(\S\ref{sec:highD} for a graphical example).  The bias is much less
pronounced in the direct estimate of ${\tilde Z}$ by the VTA owing to
smallness of the posterior probability in the extrema of the sample.
These extrema result in anomalously small value of $\log{\tilde Z}=
-289.8$ for the HMA.

\begin{figure}[thb]
  \subfigure[${\tilde M}(Y)$]{
    \includegraphics[width=0.5\textwidth]{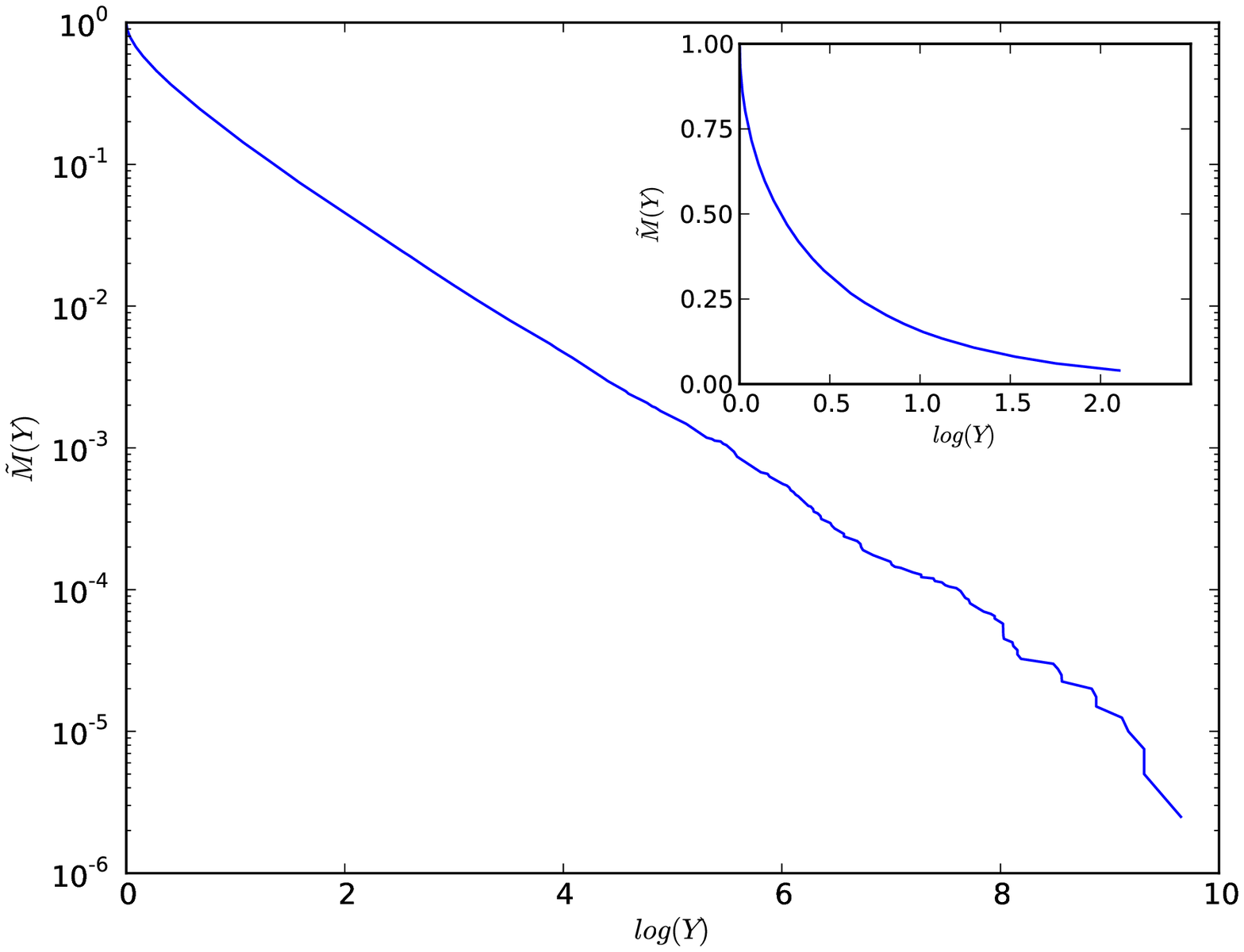}
  }
  \subfigure[${\tilde K}(Y)-Y_0$]{
    \includegraphics[width=0.5\textwidth]{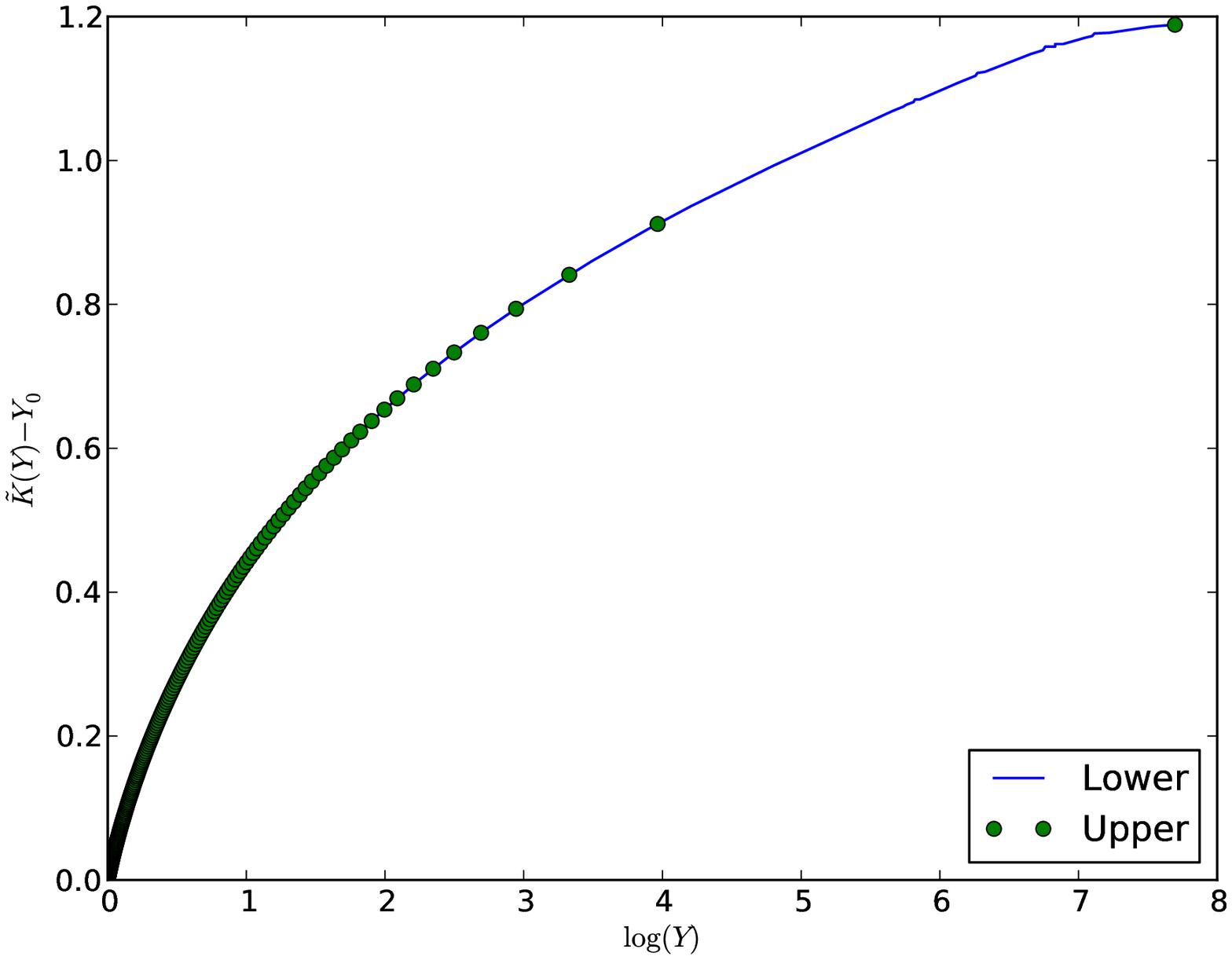}
  }
  \caption{Details of the marginal likelihood computation illustrating
    the numerical Lebesgue approach.  Panel (a) compares the run of
    measure ${\tilde M}$ function with $Y\equiv L_0/L$ computed from
    posterior simulation with $N=400000$ elements using the NLA.
    Panel (b) shows the Lebesgue \emph{quadrature} term, ${\tilde
      K}(Y)-Y_0$ from eq. \protect{\ref{eq:Kyg}}.  ${\tilde
      K}_{lower/upper}-Y_0$ are the lower and upper Riemann sums.  The
    sum ${\tilde K}$ converges as long as $M$ decreases faster than
    $1/Y$.  This illustrates the essence of the algorithm: anomalously
    small values of $L$ degrade the fidelity of ${\tilde M}$ at large
    $Y=L_0/L$ but these same values of ${\tilde M}$ make negligible
      contribution to ${\tilde K}$ and therefore, may be truncated
      from the quadrature sums.}
  \label{fig:mk}
\end{figure}

Figure \ref{fig:mk} illustrates the details of the NLA applied to this
computation.  Panel (a) plots ${\tilde M}$ from equation
(\ref{eq:measureX3}).  The run of ${\tilde M}$ with $Y$ rises rapidly
near the posterior mode and drops rapidly to zero for small likelihood
values.  The inset in this figure shows ${\tilde M}$ in linear scale.
The measure function ${\tilde M}$, and hence the integral ${\tilde
  K}$, is dominated by large values of $L$ as long as $M$ decreases
sufficiently fast (see \S\ref{sec:gauss_ex}).  Panel (b) plots the
accumulating sum defining the quadrature of ${\tilde Z}$ in equations
(\ref{eq:LUn})--(\ref{eq:Tn}), beginning with the largest values of
likelihood first.  The contribution to ${\tilde Z}$ is dominated at
the likelihood peak, corresponding to the steeply rising region of
${\tilde M}$ in Panel (a).  In other words, the samples with small
values of $L$ that degrade the HMA make a negligible contribution to
the marginal likelihood computation as long as $h_i<h_\ast$.  In
addition, NLA provides upper and lower bounds, and thereby some
warning when the estimate is poorly conditioned, e.g. owing to an
inappropriate choice for $h_\ast$.  The plot in Figure \ref{fig:mk}b
will readily reveal such failures.

\begin{figure}
  \subfigure[$-0.2<{\bar x}<1.2$]{
    \includegraphics[width=0.5\textwidth]{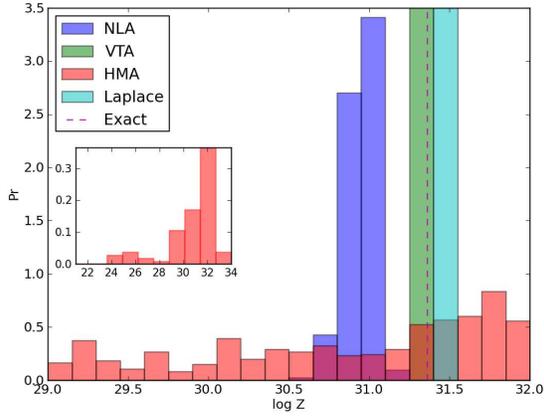}
  }
  \subfigure[$0.2<{\bar x}<0.8$]{
    \includegraphics[width=0.5\textwidth]{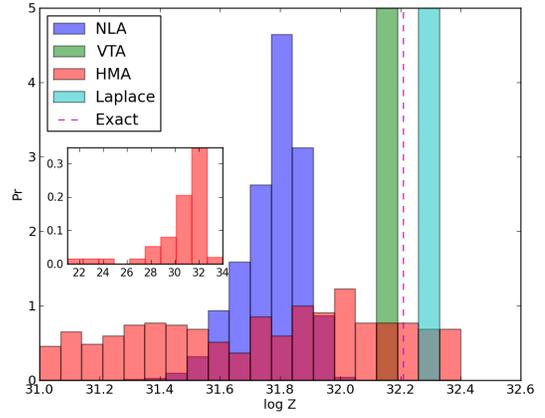}
  }
  \subfigure[$0.4<{\bar x}<0.6$]{
    \includegraphics[width=0.5\textwidth]{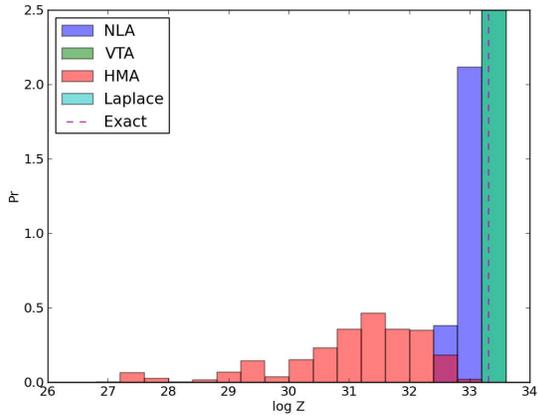}
  }
  \subfigure[$0.46<{\bar x}<0.54$]{
    \includegraphics[width=0.5\textwidth]{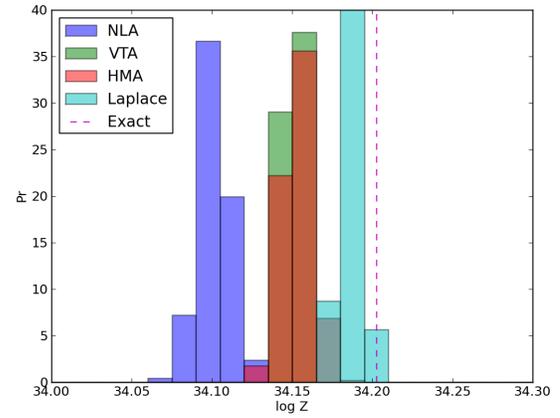}
  }
  \caption{Histogrammed distributions of ${\tilde T}$ for the NLA,
    VTA, HMA, and the Laplace approximation for 10,000 randomly
    resampled states of a converged posterior distribution of 200,000
    states.  The dashed line shows the true value computed by directly
    integrating $Z$ from eq. \protect{\ref{eq:evidence}}. Each panel
    is labeled by the range of the flat prior distribution for the
    position of the normal distribution.}
  \label{fig:subsample}
\end{figure}

A more realistic error assessment can be obtained by subsampling the
sequence $\{L_k\}$.  The CPU time for these algorithms is sufficiently
small that this procedure should be practical in general.  Consider
the following experiment: (1) the posterior is simulated by MCMC (as
described above) to obtain a chain of 400,000 states; (2) the first
half of the chain is discarded; (3) the second-half is randomly
subsampled with replacement to obtain 128 samples of 10,000 states;
(4) the marginal likelihood for each is computed using the NLA, VTA,
the Laplace approximation and the HMA (approximately 2 CPU minute in
total).  For all but the HMA, increasing the number of states
decreases the variance for each distribution; samples with 10,000
states best revealed the differences between the algorithms with a
single scale.

Figure \ref{fig:subsample} illustrates the relative performance with
different prior distributions.  Figure \ref{fig:subsample}a is the
model described at the beginning of this section; the range of the
prior distribution is much larger than the values sampled from the
posterior distribution.  The prior distributions for each successive
panel have smaller ranges as indicated.  The colors are
composited\footnote{For each color channel, value $c_1$ over $c_0$
  yields the new value $c=(1-\alpha)c_0 + \alpha c_1$.} with
$\alpha=0.5$ (e.g. HMA over VTA is brown, HMA over NLA is purple,
Laplace over HMA is blue-grey, Laplace over VTA is blue-green).  In
Panel (d), the range is within the range of values sampled by the
posterior in Panel (a).  The overall trends are as follows: 1) the HMA
has unacceptably large variance unless the domain of the prior roughly
coincides with the domain sampled by the MCMC algorithm; 2) the VTA
and Laplace approximation have the smallest variances, followed by
HMA; 3) the NLA is consistently biased below the analytic value; and
4) the VTA and Laplace approximation are closed to the expected
analytic value.  Indeed, the Laplace approximation is an ideal match
to and should do well for this simple unimodal model. In the final
panel, there are no outlier values of $L$ and the harmonic mean
approximation is comparable to the others.  These tests also
demonstrate that the same outliers that wreck the HMA have much less
affect on NLA and VTA.  Further experimentation reveals that the
results are very insensitive to the threshold value $h_\ast$.  In
fact, one needs an absurdly large value of $h_\ast$, $h_\ast>1$, to
produce failure.

\subsection{Non-nested Linear Regression Models}
\label{sec:nnlrm}

Here, we test these algorithms on the radiata pine compressive
strength data analyzed by Han and Carlin (2001) and a number of
previous authors.  We use the data tabled by Han and Carlin from
Williams (1959).  These data describe the maximum compressive strength
parallel to the grain $y_i$, the density $x_i$, and the resin-adjusted
density $z_i$ for $N = 42$ specimens.  Carlin and Chib (1995) use
these data to compare the two linear regression models:
\begin{eqnarray*}
  M = 1: y_i &=& \alpha + \beta(x_i - {\bar x}) + \epsilon_i,
  \,\,\,\quad \epsilon_i\sim {\cal N}(0, \sigma^2), \qquad i=1, \ldots, N \\
  M = 2: y_i &=& \gamma + \delta(z_i - {\bar z}) + \epsilon_i,
  \qquad \epsilon_i\sim {\cal N}(0, \tau^2), \qquad i=1, \ldots, N 
\end{eqnarray*}
with ${\cal M} = \{1, 2\}, \mathbb{\theta}_1 = \{\alpha, \beta,
\sigma^2\}^T$, and $\mathbb{\theta}_2 = \{\gamma, \delta, \tau^2\}^T$.
We follow Han and Carlin (2001) and Carlin and Chib (1995), adopting
${\cal N}\left(\{3000, 185\}^T, \mbox{Diag}\{10^6, 10^4\}\right)$
priors on $\{\alpha, \beta\}^T$ and $\{\gamma, \delta\}^T$, and
$\mbox{IG}\left(3, [2*300^2]^{-1}\right)$ priors on $\sigma^2$ and
$\tau^2$, where $\mbox{IG}(a, b)$ is the inverse gamma distribution
with density function
\[
f (v) = \frac{e^{-1/(bv)}}{\Gamma(a)b^av^{a+1}}
\]
where $v>0$ and $a,b>0$.  Han and Carlin point out these priors are
approximately centered on the least-squares solution but are otherwise
rather vague.  Using direct integration, Green and O'Hagan (1998) find
a Bayes factor of about 4862 in favor of Model 2.

\newcommand{\natop}[2]{\genfrac{}{}{0pt}{}{#1}{#2}}

\begin{table}[htb]
\centering
\caption{Marginal likelihood for non-nested linear regression models}
\label{tab:lrm}
\renewcommand{\arraystretch}{1.5}
\begin{tabular}{l|l|l|l|l} \hline
  Model & $\log Z(M=1)$& $\log Z(M=2)$ & $B_{21}$ & $\Delta\%$ \\ \hline
 NLA &  $-309.69\natop{+0.07}{-0.10}$ &  $-301.20\natop{+0.08}{-0.08}$ &  $4866\natop{+965}{-707}$ & 0.1 \\
 VTA &  $-308.30\natop{+0.02}{-0.02}$ &  $-299.83\natop{+0.02}{-0.02}$ &  $4741\natop{+189}{-186}$ & -2.5 \\
 HMA &  $-379.99\natop{+14.46}{-8.64}$ &  $-386.52\natop{+23.16}{-7.74}$ &  $0\natop{+10^{11}}{-0}$ & -100.0 \\
 Laplace &  $-306.66\natop{+0.03}{-0.03}$ &  $-298.15\natop{+0.03}{-0.04}$ &  $4974\natop{+318}{-327}$ & 2.3 \\
\hline
\end{tabular}
\end{table}

\begin{figure}[thb]
  \centering
  \includegraphics[width=0.7\textwidth]{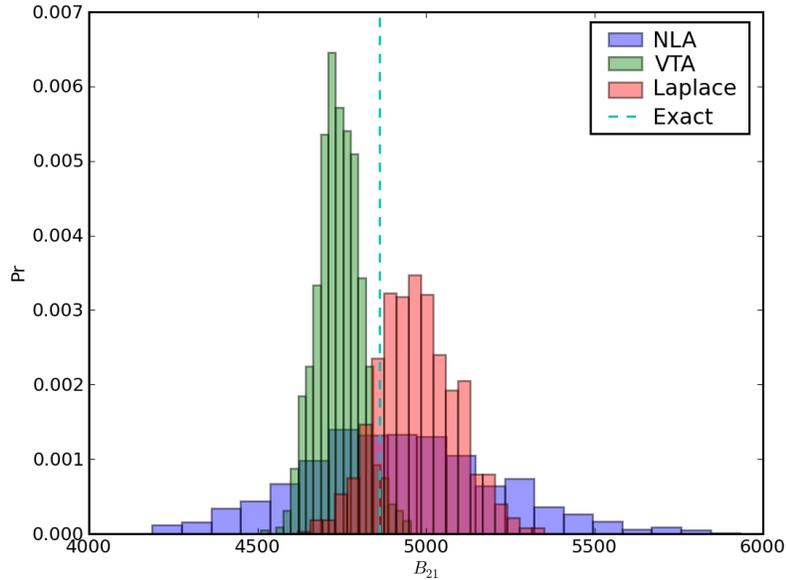}
  \caption{The histogrammed distribution of Bayes factors for the 1024
    samples using the NLA, VTA and Laplace approximation.  Although
    the variance for the NLA is larger than the VTA or Laplace
    approximation, its bias is small.}
  \label{fig:b21lr}
\end{figure}

Table \ref{tab:lrm} describes the results of applying the algorithms
from previous sections to a converged MCMC chain of 2.4 million states
for both models using the parallel tempering scheme.  The quoted value
is the median and the bounds are the $p=0.025$ and $p=0.975$ quantiles
computed from 1024 bootstrap samples of 100,000 states.  I chose
100,000 state samples to achieve 95\% confidence bounds of
approximately 10\% or smaller for both the NLA and VTA.  The second
and third columns of the table are the value of marginal likelihood
for Models 1 and 2 for each of the four models listed in the first
column.  The quoted range is the 95\% confidence bounds for each
median value from the 1024 samples.  The fourth column is the Bayes
factor for Model 2 to Model 1 and the fifth column is the relative
difference from the exact result.  The NLA, VTA and Laplace
approximation yield values within a few percent of the true value.
The VTA presents the smallest variance, followed by Laplace and then
NLA.  The HMA samples are too broadly distributed to be of use.
Figure \ref{fig:b21lr} shows the distribution of $B_{21}$ for the
samples; counter to the trend from \S\ref{sec:test}, both the VTA
and Laplace approximation are more biased than the NLA here.

\begin{figure}[thb]
  \centering
  \includegraphics[width=0.7\textwidth]{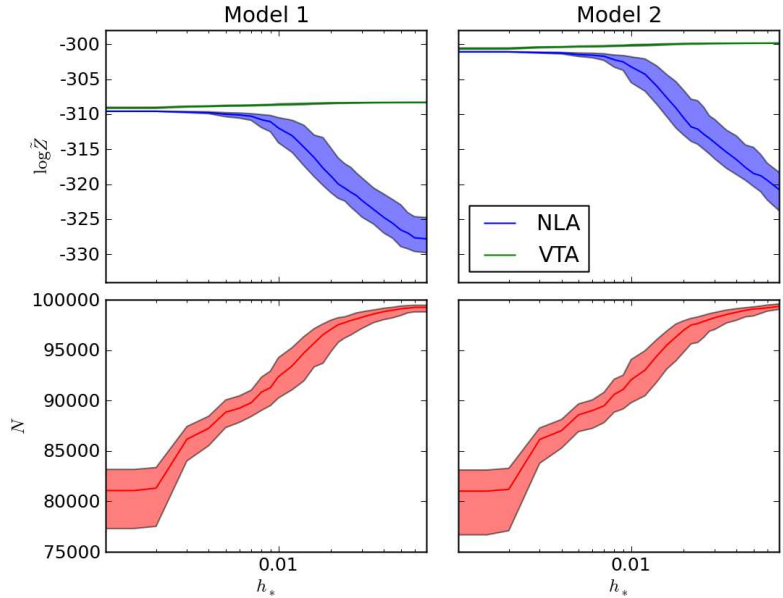}
  \caption{Comparison of the NLA and VTA as a function of $h_\ast$
    for Models 1 and 2.  The upper panel shows the run of ${\tilde Z}$
    for increasing $h_\ast$; the lower panel shows the number of
    states out of 100000 that meet the $h_\ast$ threshold criterion.
    This increases to 100000 as $h_\ast$ increases; for a threshold
    $h_\ast=0.06$, approximately 100 states are rejected.  The
    median (95\% confidence region) is shown as a solid line (shaded
    band).  The VTA 95\% confidence region is nearly indistinguishable
    from the line width!}
  \label{fig:veps}
\end{figure}

The value $h_\ast$ used to compute the NLA will vary with the problem
and the sample size. Therefore, some analysis of ${\tilde Z}$ is
required to choose an appropriate value.  As an example, Figure
\ref{fig:veps} plots the median and 95\% confidence region for the
bootstrap sampled marginal likelihood computation as a function of
$h_\ast$ for the regression problem.  The value of the VTA for the
same truncated sample is shown for reference only; truncation is not
needed for the VTA.  The values for ${\tilde Z}$ track each other
closely for $0.001\le h_\ast\le0.008$.  For $h_\ast<0.001$, there are
too few states for a reliable computation of ${\tilde Z}$. For
$h_\ast>0.008$, the NLA values are sensitive to the low-likelihood
tail, resulting in divergence with increasing $h_\ast$.

\subsection{High-dimension parameter spaces}
\label{sec:highD}

We adopt a `data-free' likelihood function for parameter vector
$\mathbf{\theta}$ with rank $k$:
\[
L(\mathbf{\theta}) = \left(2\pi\sigma^2\right)^{-k/2}
e^{-\mathbf{\theta}^2/2\sigma^2}.
\]
with $\sigma^2=\mbox{constant}$.  Further, we assume that each
parameter $\theta_j$ is normally distributed with a mean of 0 and a
variance of 1.  The resulting expression for the marginal likelihood
may be directly integrated, yielding $P(\sigma^2, k) =
\left[2\pi(1+\sigma^2)\right]^{k/2}$.

\begin{table}[htb]
\centering
\caption{Test of high-dimensional marginal likelihood}
\label{tab:highD}
\renewcommand{\arraystretch}{1.5}
\begin{tabular}{ll|ll|llll|ll}\hline
  \multicolumn{2}{c|}{Model} & \multicolumn{2}{c|}{NLA}
  & \multicolumn{4}{c|}{VTA} & \multicolumn{2}{c}{Laplace} \\ \hline
  k & Exact & \multicolumn{1}{c}{$\log{\tilde Z}$} & $\Delta\%$ &
  \multicolumn{1}{c}{$\log{\tilde Z}_{0.1}$} &
  \multicolumn{1}{c}{$\log{\tilde Z}_{0.5}$} &
  \multicolumn{1}{c}{$\log{\tilde Z}_{0.9}$} & $\Delta\%$ &
  \multicolumn{1}{c}{$\log{\tilde Z}$} & $\Delta\%$ \\ \hline
1  & -1.468 & $-1.45\natop{+0.01}{-0.01}$ & 0.7 & -1.45  &
$-1.45\natop{+0.01}{-0.01}$  & -1.45 & 0.7 &
$-1.60\natop{+0.01}{-0.01}$ & 9.0 \\
2  & -2.936 & $-2.94\natop{+0.02}{-0.04}$ & 0.5 & -2.92  &
$-2.92\natop{+0.01}{-0.01}$ & -2.92 & 0.5 &
$-3.20\natop{+0.01}{-0.01}$ & 9.0 \\
5  & -7.341 & $-7.31\natop{+0.01}{-0.01}$ & 0.4 & -6.90  &
$-7.35\natop{+0.01}{-0.01}$  & -7.48 & 0.1 &
$-7.99\natop{+0.01}{-0.01}$ & 8.7 \\
10 & -14.68 & $-14.47\natop{+0.01}{-2.79}$ & 1.4 & -14.56 &
$-14.44\natop{+0.01}{-0.01}$ & -14.34 & 1.6 &
$-16.06\natop{+0.04}{-0.03}$ & 9.4 \\
20 & -29.36 & $-29.23\natop{+0.01}{-0.01}$ & 0.4 & -29.38 &
$-29.14\natop{+0.01}{-0.01}$ & -28.91 & 0.7 &
$-32.38\natop{+0.08}{-0.07}$ & 10 \\
40 & -58.73 & $-59.51\natop{+0.23}{-0.14}$ & 1.3 & -59.69 &
$-59.01\natop{+0.19}{-0.15}$ & -58.97 & 0.9 &
$-56.59\natop{+0.01}{-0.01}$ & 8.1 \\
\hline
\end{tabular}
\end{table}

For each model of dimension $k$, we compute a Markov chain using the
Differential Evolution algorithm \citep[DE,][]{TerBraak:06}.  This
algorithm evolves an ensemble of chains with initial conditions
sampled from the prior distribution.  A proposal is computing by
randomly selecting pairs of states from the ensemble and using a
multiple of their difference; this automatically `tunes' the proposal
width.  We have further augmented this algorithm by including a
tempered simulation step \citep{Neal:96} after every 20 DE steps
\citep[see][for more details]{Weinberg:10}.

Each row describes of Table \ref{tab:highD} describes the application
of the NLA, VTA, and Laplace approximation to a model of dimension
$k$.  The MCMC simulations produce approximately 1.4 million
converged states.  Convergence is testing using the Gelman-Rubin
statistic (op. cit.). Each converged chain is resampled with
replacement to provide 1024 subsamples of $n$ states. The value
$N\in[10000,400000]$ is chosen to achieve 95\% confidence intervals
approximately 1\% of ${\tilde Z}$ or smaller.  The 95\% confidence
intervals on ${\tilde Z}$ are indicated as sub- and super-scripts.
Recall that the standard VTA determines volume spanned $\barm$ samples
and approximates the integral by multiplying the volume by the median
value of the sample.  To assess the variance inherent in this choice,
I quote the results for two other p-quantiles, $p=0.1$ and $p=0.9$.
Finally, for each algorithm, the table presents the relative error:
$\Delta\%\equiv|{\log \tilde Z}-\log Z|/|\log Z|\times100$.

Both the NLA and VTA results are very encouraging: the relative error
is within a few percent for $1\le k\le40$.  For $k=40$, I computed
${\tilde Z}$ with samples sizes of 400,000 states. Both the NLA and
VTA tend to slightly overestimate $Z$ for large $k$.  The Laplace
approximation results are disappointing for small $k$ and improve for
large $k$, but still are less precise than either the NLA or VTA.

\begin{figure}[thb]
  \centering
  \includegraphics[width=0.7\textwidth]{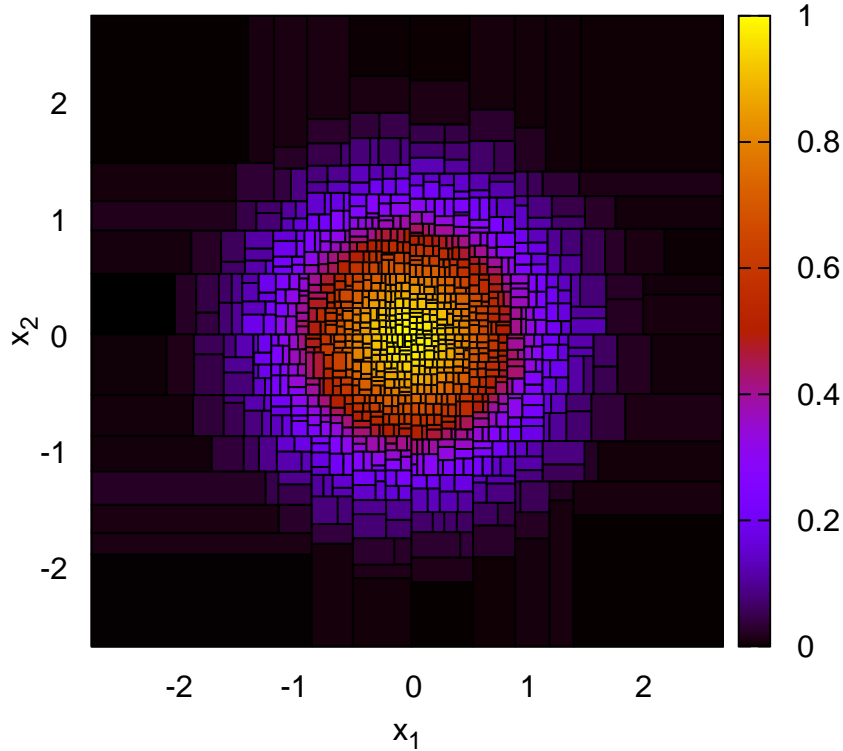}
  \caption{Two-dimensional illustration of the domain decomposition
    for the Gaussian likelihood example described in
    \S\protect{\ref{sec:highD}}.  The cells are colored according to
    posterior probability on a linear scale from 0 to $\sup\{P\}$. }
  \label{fig:k2d}
\end{figure}

Figure \ref{fig:k2d} illustrates the kd-tree construction for a single
$k=2$ sample.  Each two-dimensional cell is colored by the median
value of the posterior probability for the $\barm=32$ points in each
cell and scaled to the peak value of posterior probability $P$ for the
entire sample.  A careful by-eye examination of the cell shape reveals
a preponderance of large axis-ratio rectangles; this is a well-known
artifact of the kd-tree algorithm.  For large values of $P$, the
volume elements are small, and with a sufficiently large sample, the
gradient in $P$ across the volume are small.  For small values of $P$,
the volume elements are large, the gradients are large, and the
large-axis ratio rectangles distort the reconstructed shape of the
true posterior.  However, as described in \S\ref{sec:newevid}, the
values of $\pi(\param)L(\data|\param)\Delta V = \mbox{constant}$ for
an infinite sample, so a small number of distorted rectangles will not
compromise the end result.  Moreover, the values of
$\pi(\param)L(\data|\param)\Delta V$ at large volumes are smaller than
those at small volume for these tests, and this further decreases the
importance of the kd-tree cell-shape artifact.

\subsection{Model selection}
\label{sec:select}

\begin{figure}
  \subfigure[$x\sim {\cal N}(0.5, 0.03)$]{
    \includegraphics[width=0.5\textwidth]{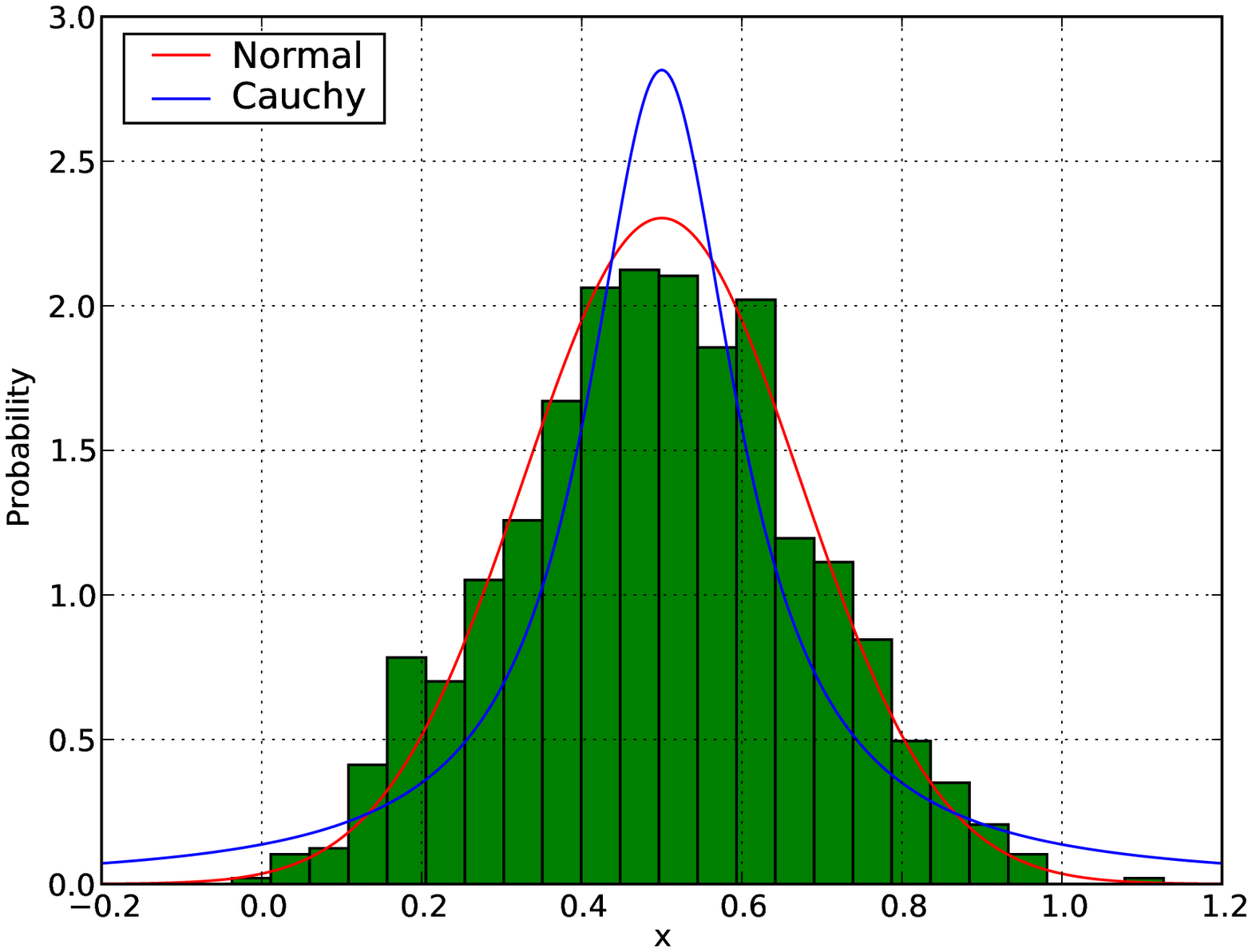}
    }
  \subfigure[$x\sim \mbox{Cauchy}(0.5, \sqrt{0.03})$]{
    \includegraphics[width=0.5\textwidth]{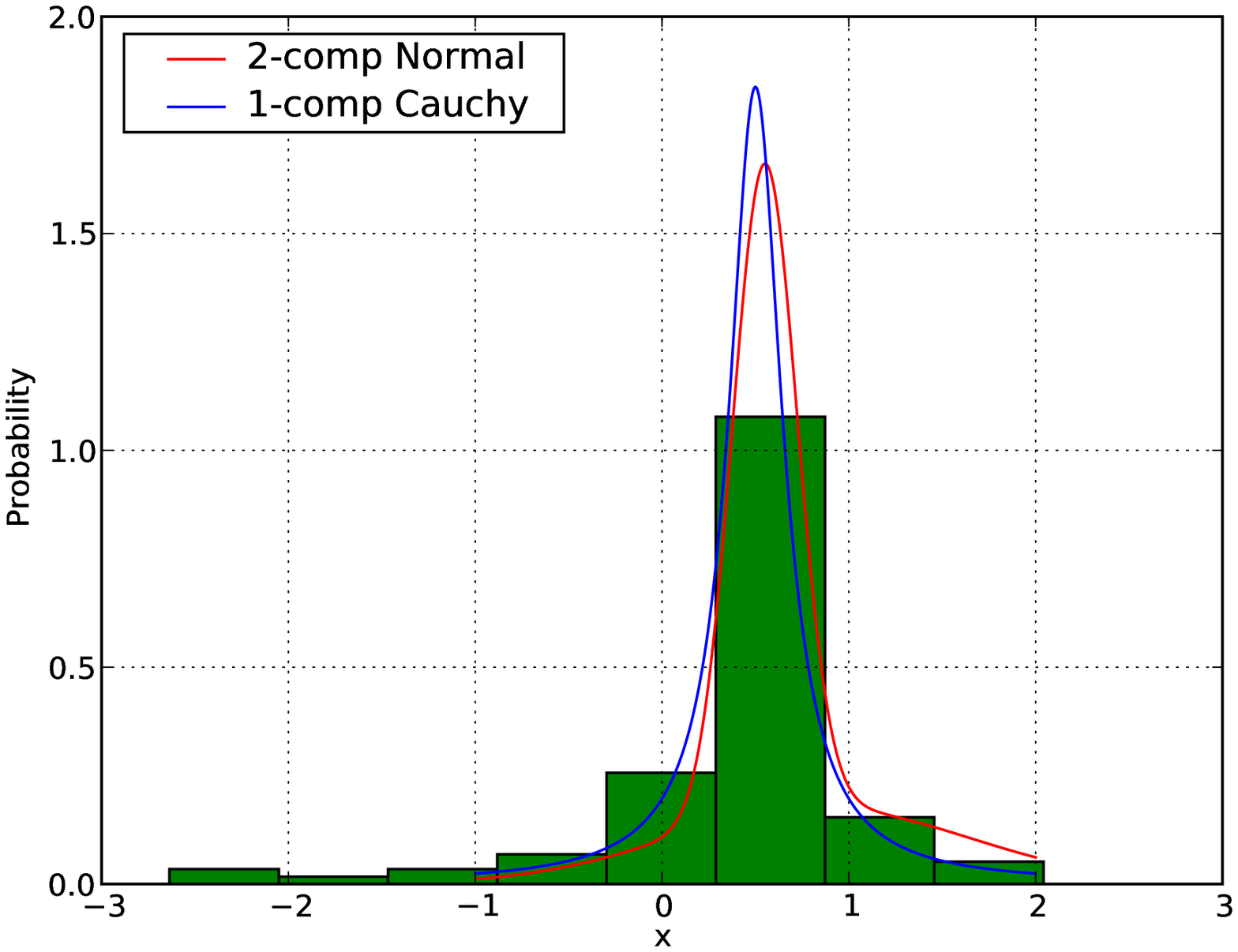}
  }
  \caption{A histogrammed distribution of the 1000 data points from
    ${\cal N}(0.5, 0.03)$ Panel (a) and 100 data points from
    ${\cal C}(0.5, \sqrt{0.03})$ Panel (b) used in these examples
    compared with the ``best fit'' Normal and Cauchy distributions
    chosen from the peak of the posterior distribution.}
  \label{fig:cauchy}
\end{figure}

\begin{figure}
  \centering
  \includegraphics[width=0.5\textwidth]{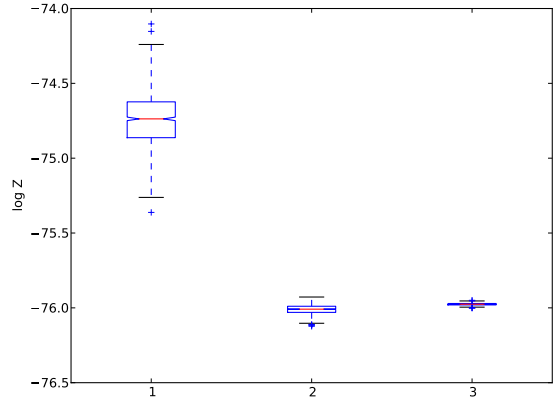}
  \caption{Box and whisker plot for the distribution of the
    $\log{\tilde Z}$ but for a sample from the Cauchy distribution
    from Fig. \protect{\ref{fig:cauchy}}.  The box shows the quantiles
    and median, the whisker shows the (10\%, 90\%) intervals, followed
    by outlying points.  The three distributions are (1) the HMA; (2)
    NLA for ${\tilde K}$ and VTA
    for ${\tilde J}$; and (3) VTA for both ${\tilde
      J}$ and ${\tilde Z}$.}
  \label{fig:subsampleC}
\end{figure}

As an example of model selection, we first compute the marginal
likelihood for the same data $x\sim{\cal N}(0.5, 0.03)$ as in the
first example of \S\ref{sec:test} but assuming a Cauchy-Lorentz
distribution,
\[
{\cal C}(a, b): P(x|a,b) = \left[\pi b\left(1+ \frac{(x -
    a)^2}{b^2}\right)\right]^{-1},
\]
as the model with unknown location $a$ and scale $b$ parameters.  For
prior distributions, we take $a\sim{\cal U}(0, 1)$ and $b\sim {\cal
  W}(0.025, 1)$ where ${\cal W}(\lambda, k)$ is the Weibull
distribution with scale parameter $\lambda$ and shape parameter $k$.
NLA yields $\log{\tilde Z} = 5.10^{+0.03}_{-0.03}$, VTA yields
$\log{\tilde Z} = 5.12^{+0.02}_{-0.01}$ and the HMA yields
$\log{\tilde Z}= 6.62^{+0.80}_{-119}$.  The data and fits are shown in
Figure \ref{fig:cauchy}a.  There should be no surprise that the true
model (with $\log Z=33.5$)is strongly preferred. Let us now repeat the
experiment using 100 data points selected from the Cauchy-Lorentz
distribution (${\tilde Z}_1$) and compare the marginal likelihood
values for a Cauchy-Lorentz distribution and a mixture of two Normal
distributions (${\tilde Z}_2$).  NLA and VTA, respectively, yield
$\log{\tilde Z}_1 = -76.3^{+0.04}_{-0.04}, -76.1^{+0.01}_{-0.01}$ and
$\log{\tilde Z}_2 = -136.9^{+0.2}_{-0.6}, -134.4^{+1.1}_{-0.6}$.  The
HMA yields $\log{\tilde Z}_1 = -75.9^{+0.01}_{-0.01}$ and $\log{\tilde
  Z}_2= -116.5^{+1.3}_{-0.9}$.  Regardless of the algorithm performing
the test, the Bayes factor reveals strong evidence in favor of the
true model.  Note from Figure \ref{fig:cauchy}b that both models are
reasonable fits ``by eye''.  However, the Bayes factor overwhelmingly
\emph{prefers} the simpler (in this case, true) model.  As expected,
the distribution of ${\tilde Z}$ for the heavy-tailed Cauchy
distribution is much better behaved (see
Fig. \ref{fig:subsampleC}). The results for NLA and VTA are consistent
and the HMA is systematically larger, but non enough to misguide a
decision.

\section{Discussion and Summary}
\label{sec:summary}

In summary, much of the general measure-theoretic underpinning of
probability and statistics naturally leads naturally to the evaluation
of expectation values.  For example, the harmonic mean approximation
\citep[HMA,][]{Newton.Raftery:94} for the marginal likelihood has
large variance and is slow to converge \citep[e.g.][]{Wolpert:02}.  On
the other hand, the use of analytic density functions for the
likelihood and prior permits us to take advantage of less general but
possibly more powerful computational techniques.  In
\S\S\ref{sec:intro}--\ref{sec:algo} we diagnose the numerical origin
of the insufficiencies of the HMA using Lebesgue integrals.  There are
two culprits: 1) the integral on the left-hand side of equation
(\ref{eq:Zdef0}) may diverge if the measure function $M(Y=L^{-1})$
from equation (\ref{eq:measureX}) decreases too slowly; and 2)
truncation error may dominate the quadrature of the left-hand side of
equation (\ref{eq:Zdef0}) unless the sample is appropriately
truncated.  Using numerical quadrature for the marginal likelihood
integral (eqns. \ref{eq:evidence} and \ref{eq:measureX}) leads to
improved algorithms: the \emph{Numerical Lebesgue Algorithm} (NLA) and
the \emph{Volume Tessellation Algorithm} (VTA).  Our proposed
algorithms are a bit more difficult to implement and have higher
computational complexity than the simple HMA, but the overall CPU time
is rather modest compared to the computational investment required to
produce the MCMC-sampled posterior distribution itself.  For a sample
of size $N$, the sorting required by NLA and VTA has computational
complexity of ${\cal O}(N\log N)$ and ${\cal O}(N\log^2 N)$,
respectively, rather than ${\cal O}(N)$ for the harmonic mean.
Nonetheless, the computational time is a fraction of second to minutes
for typical values of $10^5<N<10^8$ (see \S\ref{sec:algo}).

The geometric picture behind NLA is exactly that for Lebesgue
integration.  Consider integrating a function over a two-dimensional
domain.  In standard Riemann quadrature, one chops the domain into
rectangles and adds up their area.  The sum can be refined by
subdividing the rectangles; in the limit of infinitesimal area, the
resulting sum is the desired integral.  In the Lebesgue approach, one
draws horizontal slices through the surface and adds up the area of
the horizontal rectangles formed from the width of the slice and the
vertical distance between slices.  The sum can be refined by making
the slices thinner when needed; in the limit of slices of
infinitesimal height, the resulting sum is the desired integral.  In
the Riemann case, we multiply the box area in the domain, $dA$, by the
function height, $f$.  In the Lebesgue, we multiply the slice height
in the range, $df$, by the domain area, $A$ (see Fig. \ref{fig:geom}).
Both algorithms easily generalize to higher dimension.  For the
Lebesgue integral, the slices become level sets on the hypersurface
implied by the integrand.  Therefore the Lebesgue approach always
looks one-dimensional in the level-set value; the dimensionality $k$
is `hidden' in the area of domain (\emph{hypervolume} $A$ for $k>3$)
computed by the measure function $M(Y)$.  The level-set value for the
NLA is $Y=1/L(\data|\param)$.  Once determined, NLA applies the
trapezoidal rule to the sum over slices and compute the upper and
lower rectangle sums as bounds.  Clearly, the error control on this
algorithm might be improved by using more of the information about the
run of $A$ with $f$.

Having realized that the practical failure of the harmonic mean
approximation is a consequence of the sparsely sampled parameter-space
domain, NLA addresses the problem by determining a well-sampled subset
$\Omega_s\subset\Omega$ from the MCMC sample, ex post facto.
Restricted to this subset, $\Omega_s$, the value of the integral $J$
on the right-hand side of equation (\ref{eq:Zdef0}) is less than
unity.  We determine $\Omega_s$ by a binary space partitioning (BSP)
tree and compute $J$ from this partition.  A BSP tree recursively
partitions a the k-dimensional parameter space into convex subspaces.
The VTA is implemented with a kd-tree \citep{Cormen.etal:01} for
simplicity.  In addition, one may use VTA by itself to compute
equation (\ref{eq:evidence}) directly.

Judged by bias and variance, the test examples do not suggest a strong
preference for either the NLA or the VTA.  However, both are clearly
better than the HMA or the Laplace approximation.  Conversely, because
these algorithms exploit the additional structure implied by smooth,
well-behaved likelihood and prior distribution functions, the
algorithms developed here will be inaccurately and possibly fail
miserably for \emph{wild} density functions.  The NLA and the VTA are
not completely independent since the NLA uses the tessellation from
the VTA to estimate the integral $J$.  However, the value of the
integral $K$ tends to dominate $Z$, that is $|\log K|\gg|\log J|$, and
the contributions are easily checked.  Based on current results, I
tentatively recommend relying preferentially on VTA for the following
reasons: 1) there is no intrinsic divergence; 2) it appears to do as
well as VTA even in a high-dimensional space; and 3) there is no
truncation threshold $h_\ast$.  

Figure \ref{fig:k2d} illustrates the potential for volume artifacts
that could lead to both bias and variance.  This error source affects
both the VTA and NLA (through the computation of ${\tilde J}$) but the
affect on the NLA may be larger (\S\ref{sec:test}).  Additional
real-world testing, especially on high-dimensional multimodal
posteriors, will provide more insight.  In test problems described in
this paper, I explored the effects of varying the threshold $h_\ast$
and the kd-tree bucket size $\barm$.  These parameters interact the
sample distribution, and therefore, are likely to vary for each
problem.  I also recommend implementing both the NLA, VTA, HTM,
Laplace approximation and comparing the four for each problem.  We are
currently testing these algorithms for astronomical inference problems
too complex for a simple example; the results will be reported in
future papers.  An implementation of these algorithms will be provided
in the next release of the UMass Bayesian Inference Engine
\citep[BIE,][]{Weinberg:10}.

There are several natural algorithmic extensions and improvements not
explored here.  \S\ref{sec:intgr} describes a smoothed approximation
to the computation of $M(Y)$
(eqns. \ref{eq:nummeas}--\ref{eq:ItildeN}) rather than the step
function used in \S\ref{sec:algo}.  The direct integration of equation
(\ref{eq:evidence}) currently ignores the location of field values in
each cell volume.  At the expense of CPU time, the accuracy might be
improved by fitting the sampled points with low-order multinomials and
using the fits to derive a cubature algorithm for each cell.  In
addition, a more sophisticated tree structure may decrease the
potential for bias by providing a tessellation with ``rounder'' cells.

In conclusion, the marginal likelihood \[Z=\int
d\param\pi(\param|\model)L(\data|\param,\model)\] may be reliably
computed from a Monte Carlo posterior sample though careful attention
to the numerics.  We have demonstrated that the error in the HMA is
due to samples with very low likelihood values but significant prior
probability.  It follows that their posterior probability also very
low, and these states tend to be outliers.  On the other hand, the
converged posterior sample is a good representation of the posterior
probability density by construction.  The proposed algorithms define
the subdomain $\Omega_s\subset\Omega$ dominated by and well-sampled by
the posterior distribution and perform the integrals in equation
(\ref{eq:Zdef0}) over $\Omega_s$ rather than $\Omega$.  Although more
testing is needed, these new algorithms promise more reliable
estimates for $Z$ from an MCMC simulated posterior distribution with
more general models than previous algorithms can deliver.

\section*{Acknowledgments}

I thank Michael Lavine for thoughtful discussion and both Neal Katz
and Michael Lavine for comments on the original manuscript.  It is
also a pleasure to acknowledge the thoughtful and helpful comments of
two anonymous referees and the associate editor of the journal.  This
work was supported in part by NSF IIS Program through award 0611948
and by NASA AISR Program through award NNG06GF25G.

\end{document}